\documentclass[final,10pt,twocolumn]{elsarticle}
\usepackage[english]{babel}
\usepackage{graphicx}
\usepackage{epsfig}
\usepackage{amsmath}
\usepackage{mathtools}
\usepackage{amssymb}
\usepackage{latexsym}
\usepackage[usenames]{color}
\usepackage[spaces,hyphens]{url}

\newcommand{\dd}{\partial}

\journal{Computer Physics Communications}

\begin{document}
\begin{frontmatter}
\title{A 3+1 dimensional viscous hydrodynamic code for relativistic heavy ion collisions}
\author[lfias,lbitp]{Iu. Karpenko}
\author[lfias,litp]{P. Huovinen}
\author[lfias,litp]{M. Bleicher}

\address[lfias]{Frankfurt Institute for Advanced Studies, Ruth-Moufang-Stra{\ss}e 1, 60438 Frankfurt am Main, Germany}
\address[lbitp]{Bogolyubov Institute for Theoretical Physics, 14-b, Metrolohichna str., 03680 Kiev, Ukraine}
\address[litp]{Institute for Theoretical Physics, Johann Wolfgang Goethe Universit\"at, Max-von-Laue-Str. 1, 60438 Frankfurt am Main, Germany}
\begin{abstract}
 We describe the details of 3+1 dimensional relativistic hydrodynamic code for the simulations of quark-gluon/hadron matter expansion in ultra-relativistic heavy ion collisions. The code solves the equations of relativistic viscous hydrodynamics in the Israel-Stewart framework. With the help of ideal-viscous splitting, we keep the ability to solve the equations of ideal hydrodynamics in the limit of zero viscosities using a Godunov-type algorithm. Milne coordinates are used to treat the predominant expansion in longitudinal (beam) direction effectively. The results are successfully tested against known analytical relativistic inviscid and viscous solutions, as well as against existing 2+1D relativistic viscous code.
\end{abstract}
\end{frontmatter}

\section{Introduction}

Relativistic fluid dynamics has been applied to various high energy
phenomena in astrophysics, nuclear and hadron physics, from collision
of galaxies down to the evolution of femtometer-size droplets of dense
matter created in ultra-relativistic heavy ion collisions. In
astrophysics typical applications of relativistic fluid dynamics are
collapse of massive stars, formation of and flow around black holes,
collisions of neutron stars and passage of relativistic jets through
intergalactic matter~\cite{einstein,Marti}. On earth relativistic
flows appear in ultrarelativistic heavy-ion collisions, where the
formed matter depicts collective behaviour. Especially the
anisotropies of the final particle distribution were described so well
using ideal fluid dynamics, that the matter was called almost perfect
fluid with the lowest possible viscosity. The determination of the
dissipative properties of this matter has became one of the major
goals of heavy-ion physics, and requires sophisticated fluid
dynamical calculations.

The equations of motion of relativistic fluid dynamics are notoriusly
difficult to solve. Except in very idealised situations, no analytic
solutions exist, and the equations must be solved numerically. Several
groups have developed several codes for fluid dynamical modeling of
heavy-ion collisions~\cite{Hirano, KolbHeinz, Kodama, BassNonaka, VISH2p1, Romatschke, Muronga, PrattVredevoogd, PetersenBleicher, Molnar, Schenke, WernerKarpenko, KarpenkoSinyukov, HolopainenEskola, Bozek, BleicherGPU, NonakaTI, N-H, DelZanna}\footnote{We apologise to our
  colleagues whose work we forgot to mention.}, but many of these
codes assume boost-invariant longitudinal expansion~\cite{Bjorken}
and/or zero net baryon density in the entire system. Neither of these
assumptions is a good approximation in collisions at the Beam Energy
Scan energies ($\sqrt{s_\mathrm{NN}} = 6.3$--39 GeV) at BNL RHIC
(Relativistic Heavy-Ion Collider) nor in collisions in the forthcoming
experiments at FAIR or NICA. We have therefore developed a new code
where both of these assumptions have been relaxed. In this paper we
present the results of test simulations of this code.

High-Resolution Shock-Capturing (HRSC) algorithms are particularly
suitable for solving the equations of relativistic fluid dynamics, and
are applied for a wide variety of problems~\cite{Marti}. HRSC
algorithms are designed to treat discontinuous shock configurations in
hydrodynamic solution, or shock waves. The methods usually incorporate
higher-order schemes which minimize numerical errors.  Most of HRSC
algorithms are formulated in conservative form, where the time
evolution of cell averaged quantities is governed by numerical fluxes
evaluated at cell boundaries. The conservative form ensures that the
total energy and momentum in the system is conserved during the time
evolution. A sub-family of HRSC algorithms are Godunov-type
algorithms, which are based on exact or approximate solutions of the
Riemann problem at the cell boundaries in order to compute
time-averaged fluxes through it.

%In this paper we present the results of test simulations of a
%hydrodynamic code for heavy-ion physics. 
Our code is
based on the Godunov-type relativistic Harten-Lax-van~Leer-Einfeldt (HLLE) approximate Riemann solver \cite{HLLE, SchneiderRischke}. This particular choice of the approximate Riemann solver is motivated by its simplicity, reliability, and stability for the simulations related to the physics of ultra-relativistic heavy ion collisions. The Riemann problem is formulated for an inviscid fluid, where shock wave solutions are allowed. Basing on the algorithms established for inviscid fluid, we aim to study the evolution of nearly ideal fluid (fluids with close-to-minimal viscosity) like the one presumably created in ultrarelativistic heavy ion collisions. To do this, we employ additional methods to solve the equations of relativistic viscous hydrodynamics in the Israel-Stewart framework \cite{IsraelStewart}, keeping the ability to solve the equations of ideal hydrodynamics in the limit of zero shear and bulk viscosities. The use of an (approximate) Riemann solver makes it possible to treat the highly inhomogeneous matter configurations emerging from event-by-event initial conditions as employed in the most recent studies of heavy ion collisions.

The present hydrodynamic code is already being used as a part of EPOS3 event generator for ultra-relativistic heavy ion collisions \cite{epos3} and as a part of hydrodynamic+cascade model \cite{energyScan} in a studies focused on Beam Energy Scan (BES) project at the BNL Relativistic Heavy Ion Collider (RHIC).

The article is organized as follows. In Sec.~\ref{secEquations} the formalism is presented, Sec.~\ref{secNumerical} provides the details of the numerical implementation. Sec.~\ref{secTests} is devoted to the description and results of test simulations, including a comparison for the physical setup for the matter expansion in relativistic A+A collisions, and we summarize in Sec.~\ref{secConclusions}. 

\section{Equations}\label{secEquations}
Throughout this work natural units are employed, i.e. the speed of light in vacuum $c=1$, the Boltzmann constant $k_B=1$ and the Planck constant $\hbar=1$.

The equations of relativistic (viscous) hydrodynamics follow from the laws of energy-momentum and charge conservation:
\begin{align}
 \dd_{\nu} T^{\mu\nu}&=0, \nonumber\\
 \dd_{\nu} N^\nu_c &=0, \label{energyConserv}
\end{align}
with $T^{\mu\nu}$ being the energy-momentum tensor and $N^\nu_c$ the charge current, index $c$ enumerates the conserved charges if there are multiple conserved charges in the system.

The Landau definition of flow velocity $u^\mu$ (Landau frame) as a flow of energy \cite{LandauBook} is adopted, i.e. $\epsilon u^\mu=T^\mu_\nu u^\nu$. In this frame, the energy-momentum tensor for a viscous fluid can be decomposed as:
$$T^{\mu\nu}=\epsilon u^\mu u^\nu -(p+\Pi)\Delta^{\mu\nu}+\pi^{\mu\nu},$$
$$N^\mu_c=n_c u^\mu + V^\mu_c,$$
where
\begin{itemize}
\item $\epsilon$ and $p$ are energy density in fluid rest frame and equilibrium pressure, respectively,
\item $\Delta^{\mu\nu}=g^{\mu\nu}-u^\mu u^\nu$ is the projector orthogonal to $u^\mu$,
\item $\pi^{\mu\nu}$ and $\Pi$ are the shear stress tensor and bulk pressure,
\item $V^\mu_c$ are charge diffusion currents.
\end{itemize}

The hydrodynamic equations are closed with the equation of state (EoS) $p=p(\epsilon,n_c)$, which has to be supplied from some external model.

In the Israel-Stewart framework of relativistic viscous hydrodynamics \cite{IsraelStewart} the shear stress tensor and bulk pressure are independent dynamical variables. Recent studies \cite{DenicolRischke} show that there can be infinitely many choices for the explicit form and coefficients in equations of motion for $\pi^{\mu\nu}$ and $\Pi$. In the present work the following choice for the equations of motion for the shear stress tensor and bulk pressure is used, where we neglect vorticity terms:
\begin{subequations}
\begin{align}
<u^\gamma \dd_{;\gamma} \pi^{\mu\nu}>&=-\frac{\pi^{\mu\nu}-\pi_\text{NS}^{\mu\nu}}{\tau_\pi}-\frac 4 3 \pi^{\mu\nu}\dd_{;\gamma}u^\gamma, \label{evolutionShear}\\
u^\gamma \dd_{;\gamma} \Pi&=-\frac{\Pi-\Pi_\text{NS}}{\tau_\Pi}-\frac 4 3 \Pi\dd_{;\gamma}u^\gamma, \label{evolutionBulk}
\end{align}
\end{subequations}
and where $\dd_{;\mu}$ denotes a covariant derivative.
This choice has already been widely used in recent simulations of nucleus-nucleus collisions at relativistic energies. For the purpose of the tests and our current applications we do not include the baryon/electric charge diffusion, i.e. $V_c^\mu=0$. Angle brackets in (\ref{evolutionShear}) are defined as:
$$<A^{\mu\nu}>=(\frac 1 2 \Delta^\mu_\alpha \Delta^\nu_\beta+\frac 1 2 \Delta^\nu_\alpha \Delta^\mu_\beta - \frac 1 3 \Delta^{\mu\nu}\Delta_{\alpha\beta})A^{\alpha\beta},$$
and denote the symmetric, traceless and orthogonal to $u^\mu$ part of $A^{\mu\nu}$.
\begin{align}
\pi^{\mu\nu}_\text{NS}&=\eta(\Delta^{\mu\lambda}\dd_{;\lambda}u^\nu+\Delta^{\nu\lambda}\dd_{;\lambda}u^\mu)-\frac 2 3 \eta\Delta^{\mu\nu}\dd_{;\lambda}u^\lambda, \nonumber\\
\Pi_\text{NS}&=-\zeta \dd_{;\lambda}u^\lambda,
\end{align}
are the values of shear stress tensor and bulk pressure in limiting Navier-Stokes case.

For hydrodynamic simulations related to the physics of ultrarelativistic heavy ion collisions Milne coordinates for the $t-z$ plane in spacetime ($z$ being the collision axis) are chosen. The new coordinates are expressed in terms of Minkowski coordinates $\{t,x,y,z\}$ as $\tau=\sqrt{t^2-z^2}$, $\eta=\frac 1 2 {\rm ln}((t+z)/(t-z))$, while the definitions of $x$ and $y$ coordinates are unchanged.

The form of hydrodynamic equations in arbitrary coordinate systems is:
\begin{align}
 \dd_{;\nu} T^{\mu\nu}&=\dd_\nu T^{\mu\nu}+\Gamma^\mu_{\nu\lambda}T^{\nu\lambda}+\Gamma^\nu_{\nu\lambda}T^{\mu\lambda}=0,\label{hydro-eqns} \\
 \dd_{;\nu} N^\nu_c &= \dd_\nu N^\nu_c + \Gamma^\nu_{\nu\lambda}N^\lambda_c = 0, \nonumber
\end{align}
where %$\dd_{;\nu}$ denotes a covariant derivative and 
$\Gamma^\mu_{\nu\lambda}$ are affine connections or Christoffel symbols.

We choose West coast convention $(+,-,-,-)$ for metric tensor in Minkowski spacetime, so in Milne coordinates the invariant interval is: $ds^2=dt^2-dx^2-dy^2-\tau^2 d\eta^2$, and the metric tensor is $$g^{\mu\nu}=diag(1,-1,-1,-1/\tau^2)$$

Although spacetime is still flat, there are nontrivial Christoffel symbols, the nonzero components being:
$$\Gamma^\eta_{\tau\eta}=\Gamma^\eta_{\eta\tau}=1/\tau,\quad \Gamma^\tau_{\eta\eta}=\tau,$$
which leads to the following explicit form of hydrodynamic equations:
\begin{align}\label{eqns0}
&\dd_\nu T^{\tau\nu}+\tau T^{\eta\eta}+\frac 1 \tau T^{\tau\tau}=0, \nonumber\\
&\dd_\nu T^{x\nu}+\frac 1 \tau T^{x\tau}=0, \nonumber\\
&\dd_\nu T^{y\nu}+\frac 1 \tau T^{y\tau}=0, \\
&\dd_\nu T^{\eta\nu}+\frac 3 \tau T^{\eta\tau}=0, \nonumber\\
&\dd_\nu N^\nu_c+\frac{1}{\tau} N^\tau_c=0. \nonumber
\end{align}

In Milne coordinates, $T^{\mu\nu}$ and $N^\nu$ keep the same structure, however the velocities are expressed through the longitudinal/transverse rapidities in the Cartesian frame as:
\begin{multline}
u^\mu=\left\{u^\tau,u^x,u^y,u^\eta\right\}=
\left({\rm cosh}(\eta_f-\eta){\rm cosh}\eta_T,\right.\\
 \left.{\rm sinh}\eta_T\{\cos\phi,\sin\phi\},
\frac 1 \tau {\rm sinh}(\eta_f-\eta){\rm cosh}\eta_T\right) \label{umu}
\end{multline}
where $\eta_f=0.5{\rm ln}(1+v_z)/(1-v_z)$ is longitudinal flow rapidity and $\eta_T={\rm arctanh}(v_T/\sqrt{1-v_z^2})$ is transverse flow rapidity.
From the equation above one can see that $u^\eta=0$ when $\eta_f=\eta$, which means that $u^\eta=0$ corresponds to scaling Bjorken flow in Cartesian coordinates, $v_z=z/t$. Thus, Milne coordinates naturally describe the expansion along $z$ axis from a point-like source.

As one can see, almost all source terms in (\ref{eqns0}) are proportional to $1/\tau$, which makes them dominant for the hydrodynamic evolution at small $\tau$. This is natural when one remembers that the gradient of the longitudinal scaling flow is inversely proportional to $t$ in the Cartesian frame. The accurate numerical solution would eventually require to apply a higher order numerical time integration scheme. We circumvent this by redefining the variables in Milne coordinates as:
\begin{align}
T^{\mu\nu}&=\tilde{T}^{\mu\nu},\ \mu,\nu\neq\eta \\
T^{\mu\eta}&=\tilde{T}^{\mu\eta}/\tau,\ \mu\neq\eta\\
T^{\eta\eta}&= \tilde{T}^{\eta\eta}/\tau^2 \\
N_c^\eta &= \tilde{N_c}^\eta/\tau  \label{tildeTransform}
\end{align}
Rewriting the equations for $\tau T^{\mu\nu}$:
 \begin{align}\label{eqns1}
 &\tilde\dd_\nu(\tau \tilde T^{\tau\nu})+\frac{1}{\tau}(\tau \tilde T^{\eta\eta})=0, \nonumber\\
 &\tilde\dd_\nu (\tau \tilde T^{x\nu}) = 0, \nonumber\\
 &\tilde\dd_\nu (\tau \tilde T^{y\nu}) = 0,\\
 &\tilde\dd_\nu(\tau \tilde T^{\eta\nu}) + \frac{1}{\tau} (\tau \tilde T^{\eta\tau}) = 0, \nonumber\\
 &\tilde\dd_\nu(\tau \tilde N_c^\nu)=0, \nonumber
 \end{align}
with
$$\tilde\dd_\mu\equiv\{ \dd/\dd\tau, \dd/\dd x, \dd/\dd y, (1/\tau)\dd/\dd\eta \},$$
all the components of $\tilde{T}^{\mu\nu}$ have the same units as well as $\tilde\dd_\mu$ [1/length]. The actual conserved variables used in the code are then $Q=\{ \tau\tilde T^{\mu\tau}, \tau\tilde N^\tau \}$, fluxes are $\{ \tau\tilde T^{ij}, \tau\tilde N^i \}$, so that
$\tilde T^{\eta\eta}=(\epsilon+p)\tilde u^\eta \tilde u^\eta+p$ and $\tilde u^\eta$ does not include the factor $1/\tau$ (cf. Eq. \ref{umu}).
Then Eq. \ref{eqns1} provides the explicit form of the energy-momentum and charge conservation equations which are solved numerically.

In the same way as it was done for energy-momentum conservation equations, we separate the factors $1/\tau$ from $\pi^{\mu\nu}$ as follows: $\pi^{\mu\eta}=\tilde \pi^{\mu\eta}/\tau$,\ $\pi^{\eta\eta}=\tilde \pi^{\eta\eta}/\tau^2$, as well as $u^\eta=\tilde u^\eta/\tau$ and $\dd_\eta\rightarrow(1/\tau)\dd_\eta$. Then we rewrite (\ref{evolutionShear},\ref{evolutionBulk}) in terms of tilded variables:
\begin{align}
 \tilde \gamma\left(\dd_\tau+\tilde v^i\tilde\dd_i\right)\tilde\pi^{\mu\nu}&=-\frac{\tilde\pi^{\mu\nu}-\tilde\pi_\text{NS}^{\mu\nu}}{\tau_\pi}+I_\pi^{\mu\nu} \label{evolShearFinal}\\
 \tilde \gamma\left(\dd_\tau+\tilde v^i\tilde\dd_i\right)\Pi&=-\frac{\Pi-\Pi_\text{NS}}{\tau_\Pi}+I_\Pi \label{evolBulkFinal}
\end{align}
and solve the above equations numerically. Here $\tilde\gamma=u^0$ and $\tilde{v}^i=\tilde{u}^i/u^0$ ($i=x,y,\eta$) are the components of 3-velocity. The additional source terms are:
\begin{align}
 I_\pi^{\mu\nu}&=-\frac 4 3 \tilde \pi^{\mu\nu}\tilde\dd_{;\gamma}\tilde u^\gamma-[\tilde u^\nu \tilde \pi^{\mu\beta}+\tilde u^\mu \tilde \pi^{\nu\beta}]\tilde u^\lambda\tilde\dd_{;\lambda} \tilde u_\beta-I^{\mu\nu}_{\pi,G}, \label{Ipi}\\
 I_\Pi&=-\frac 4 3 \Pi\,\tilde\dd_{;\gamma}\tilde u^\gamma,
\end{align}
where in a given coordinate system all covariant derivatives of the four-velocity are equal to ordinary derivatives, except for:
\begin{align}
\tilde\dd_{;\eta}u^\tau&=\tilde\dd_\eta u^\tau+\tilde{u}^\eta/\tau,\;\;\; \tilde\dd_{;\eta}\tilde{u}^\eta=\tilde\dd_{\eta}\tilde{u}^\eta+u^\tau/\tau,
\end{align}
so that $\tilde\dd_{;\gamma}\tilde{u}^\gamma=\tilde\dd_{\gamma}\tilde{u}^\gamma+u^\tau/\tau$.
Also, $I^{\mu\nu}_{\pi,G}$ denote geometrical source terms (coming from Christoffel symbols):
\begin{align}
 & I^{\tau\tau}_{\pi,G}=2\tilde  u^\eta \tilde \pi^{\tau\eta}/\tau  &  I^{\tau x}_{\pi,G}=\tilde u^\eta \tilde \pi^{\eta x}/\tau  \nonumber\\
 & I^{\tau y}_{\pi,G}=\tilde u^\eta \tilde \pi^{\eta y}/\tau  &  I^{\tau\eta}_{\pi,G}=\tilde u^\eta (\tilde \pi^{\tau\tau}+\tilde \pi^{\eta\eta})/\tau  \nonumber\\
 & I^{\eta x}_{\pi,G}=\tilde u^\eta \tilde \pi^{\tau x}/\tau  &  I^{\eta y}_{\pi,G}=\tilde u^\eta \tilde \pi^{\tau y}/\tau  \nonumber\\
 & I^{\eta\eta}_{\pi,G}=2 \tilde u^\eta \tilde \pi^{\tau\eta}/\tau  &  I^{xx}_{\pi,G}=I^{xy}_{\pi,G}=I^{yy}_{\pi,G}=0  \nonumber\\
\end{align}
Most of the tests presented below, as well as heavy-ion related simulations, are performed in Milne coordinates. However, to perform shock tube test we use a version of the code which works in Cartesian coordinates. In the latter case we solve the original hydrodynamic equations (\ref{energyConserv}), as well as $I^{\mu\nu}_{\pi,G}=0$ in (\ref{Ipi}). As a result, transformations (\ref{tildeTransform}) and (\ref{eqns1}) and tilde notation in general are not used, and $z$ coordinate stands for the third direction in space.

\section{Numerical implementation}\label{secNumerical}
Let us rewrite Eq. \ref{energyConserv} in a form of evolution equations in Minkowski spacetime for simplicity:
\begin{align}
\frac{\dd Q^\mu}{\dd t} + \frac{\dd F^{\mu i}}{\dd x_i}=0, \label{conservativeForm} \\
\frac{\dd N^0}{\dd t} + \frac{\dd N^{i}}{\dd x_i}=0,  \label{conservativeForm2}
\end{align}
where index $i$ denotes spatial dimensions. $T^{0\mu}\equiv Q^\mu$ and $N_0$ are conventionally called conserved quantities, if the fluxes at the spatial boundaries of the system vanish or compensate each other, then $\int Q^\mu(t,x) d^3 x$ is conserved. Also, $T^{\mu i}\equiv F^{\mu i}$ and $N^i$ are flux terms.

We use a finite volume method to solve hydrodynamic equations (\ref{conservativeForm}),(\ref{conservativeForm2}). In this method, one works in terms of the averaged values of $T^{\tau\mu}$ in mesh $i$, which in one dimension reads:
$$\vec Q^n_i=\frac{1}{\Delta x}\int_{x_i-\Delta x/2}^{x_i+\Delta x/2} \{ T^{0 \mu}(t_i,x), N_c^0(t_i,x) \} dx,$$
and time-averaged fluxes through left and right facets of the mesh:
$$F^n_{i\pm 1/2}=\frac{1}{\Delta t}\int\limits_{t_n}^{t_{n}+\Delta t} \{ T^{x\mu}(t,x_i\pm \frac{\Delta x}{2}), N^x(t,x_i\pm \frac{\Delta x}{2}) \} dt.$$
Then, integrating the conservation laws (\ref{conservativeForm}) within $[t_n,t_n+\Delta t]$ and $[x_i-\Delta x/2,x_i+\Delta x/2]$, one gets the exact relation between the conserved quantities and the fluxes:
\begin{multline}\label{num1}
\frac{1}{\Delta t}(Q^{n+1}_{i}-Q^n_{i})+\frac{1}{\Delta x_i}(F^n_{i+1/2}-F^n_{i-1/2})=0,
\end{multline}
which can be used to propagate $Q^n$ to the next timestep. The idea of the Godunov method \cite{Godunov} is to take a piecewise uniform distribution of $T^{0\mu},N^0_c$ on a mesh and to provide an estimate for $F^n_{i\pm 1/2}$ based on exact or approximate solution of the Riemann problem at $x=x_i\pm 1/2$ with initial left and right state parameters $Q^n_i$ and $Q^n_{i+1}$, respectively. In the next timestep, the wave structure from a Riemann problem at previous timestep is completely discarded and piecewise uniform distributions for $Q^{n+1}$ are used again. One can estimate the criterion of stability for such schemes from the Courant-Friedrichs-Lewy condition \cite{CFLpaper}, which is a necessary condition for numerical scheme to be stable. For the Godunov scheme, the criterion is $2b_\text{max}\Delta t<\Delta x$, where $b_\text{max}$ is a maximal value of the signal velocity. To be on the safe side, we assume $b_\text{max}=c=1$ and use $2\Delta t\le\Delta x$.

In Milne coordinates the definition of conserved quantities and fluxes are modified as deduced from the transformed energy-momentum conservation equations (\ref{eqns1}):
$$\vec Q^n_i=\frac{1}{\Delta x}\int_{x_i-\Delta x/2}^{x_i+\Delta x/2} \{ \tau_i T^{\tau \mu}(\tau_i,x), \tau_i N_c^\tau(\tau_i,x) \} dx,$$
\begin{multline}
F^n_{i\pm 1/2}=\frac{1}{\Delta \tau}\int_{\tau_n}^{\tau_{n}+\Delta\tau} \{ \tau T^{x\mu}(\tau,x_i\pm \Delta x/2),\\ \tau N^x(\tau,x_i\pm \Delta x/2) \} d\tau, \nonumber
\end{multline}
as well as now there are nonzero source terms in (\ref{num1}). For the case of a viscous fluid one can decompose the conserved quantities and fluxes into their ideal and viscous parts:
\begin{multline}\label{num2}
\frac{1}{\Delta t}(Q^{n+1}_{\text{id},i}+\delta Q^{n+1}_i-Q^n_{\text{id},i}-\delta Q^n_i)+ \\
+\frac{1}{\Delta x_i}(\Delta F_\text{id}+\Delta\delta F)+S_{\text{id},i}+\delta S_i=0
\end{multline}
where $\Delta F=F_{i+1/2}-F_{i-1/2}$, and $\delta Q, \delta F, \delta S$ denote viscous corrections to conserved quantities, fluxes and source terms respectively.

Then, the effects of ideal and viscous fluxes/sources in Eq. (\ref{num2})  can be accounted for separately, in the same way as it is done in \cite{TakamotoInutsuka}:
\begin{equation}
\frac{1}{\Delta t}(Q^{*n+1}_{\text{id},i}-Q^n_{\text{id},i})+\frac{1}{\Delta x_i}\Delta F_\text{id}+S_{\text{id},i}=0\label{num1ideal}
\end{equation}
\begin{multline}
\frac{1}{\Delta t}(Q^{n+1}_{\text{id},i}+\delta Q^{n+1}_i-Q^{*n+1}_{\text{id},i}-\delta Q^n_i)+ \\
+\frac{1}{\Delta x_i}\Delta\delta F+\delta S_i=0 \label{num1visc}
\end{multline}
Note that there are only ideal quantities in (\ref{num1ideal}), whereas (\ref{num1visc}) describes viscous corrections to the evolution.

The full solution (\ref{num2}) for one timestep then proceeds in the substeps:\\
\textbf{Substep 1)} $Q^{*n+1}_\text{id}$ is obtained by evolving only the ideal part of the energy-momentum tensor, Eq. \ref{num1ideal} over the full timestep $\Delta t$ using Godunov-type method.\\
\textbf{Substep 2)} The Israel-Stewart equations Eq. \ref{evolShearFinal},\ref{evolBulkFinal} are solved to propagate $\pi^{\mu\nu}$ and $\Pi$ for the next timestep.
Here one has to know the values of shear/bulk terms in the Navier-Stokes limit, $\pi^{\mu\nu}_\text{NS}, \Pi_\text{NS}$, which depend on velocity gradients. We calculate $\pi^{\mu\nu}_\text{NS}, \Pi_\text{NS}$ at $n+1/2$ (half-step) using $s=s^{*(n+1/2)}$ and
 $$\dd_\tau u^\mu=((u^\mu)^{*n+1}_i-u^{\mu,n}_i)/\Delta\tau$$
 $$\dd_{x_i} u^\mu=((u^\mu)^{*(n+1/2)}_{i+1}-(u^\mu)^{*(n+1/2)}_{i-1})/(2\Delta x_i)$$ where central differences are used for second order of accuracy. The asterisk (*) denotes the values obtained from substep 1 (updated with only ideal fluxes/sources).\\
\textbf{Substep 3)} $Q^{n+1}_{\text{id},i}+\delta Q^{n+1}_i=Q^{n+1}_\text{full}$ is obtained by evolving Eq. (\ref{num1visc}) over the full timestep $\Delta t$ with viscous fluxes/sources only.
The initial condition for this substep is $Q_\text{ini}=Q^{*n+1}_\text{id}+\delta Q^n$, the first term obtained from the solution of substep 1.\\
To update $Q_\text{full}$ according to Eq. (\ref{num1visc}), we use edge/half-step values of flux/source terms $\delta F^{n+1/2}_{i\pm 1/2}$, $\delta S^{n+1/2}$, saved at substep 2.

Note that for the splitting itself, Eq. \ref{num1ideal},\ref{num1visc}, no assumption of the smallness of the viscous corrections is needed. However the assumption becomes necessary when we calculate the fluxes in the evolution equations. For example, when one calculates $F_\text{id}$ for the ideal substep one assumes that the Godunov method works well, which is proven to be the case for hydrodynamics of inviscid fluid. Thus, we can apply the scheme for nearly perfect fluids, keeping in mind that viscosity should only introduce (small) corrections to the evolution.

In what follows we describe substeps 1 and 2 in detail, whereas the application of Eq. (\ref{num1visc}) for substep 3 is straightforward.

\subsection{Ideal substep}

For the Godunov-type method employed in substep 1, we use an approximate solution to the Riemann problem constructed with the relativistic extension of the HLLE solver. Below we provide the main points of the method, whereas for a detailed description the reader is referred to \cite{SchneiderRischke}.

\begin{figure}
\begin{center}
\includegraphics[width=0.4\textwidth]{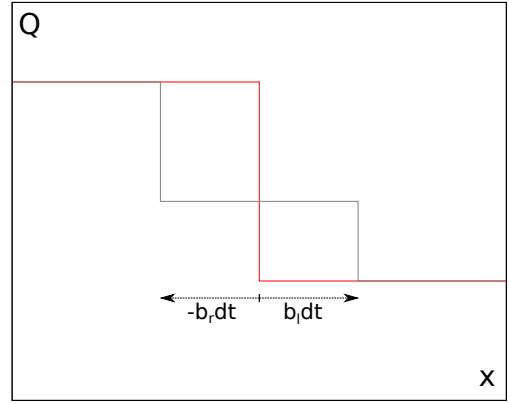}
\end{center}
\caption{(Color online) Evolution of Riemann problem in HLLE approximation. Red line represents the initial discontinuity, grey line represents the intermediate state in HLLE approximation.}\label{figHLLE}
\end{figure}

In the HLLE method, the evolution of initial discontinuity (Riemann problem) between left $Q_l=Q_i$ and right $Q_r=Q_{i+1}$ states is approximated by a single uniform intermediate state bounded by two shock waves propagating to the left and to the right from the initial discontinuity, as seen in Fig. \ref{figHLLE}. 
Within this approximation, by integrating the hydrodynamic equations over $[b_l\Delta t, b_r\Delta t]$ and $[t^n, t^n+\Delta t]$ one can derive the properties of the intermediate state with the algebraic relations:
\begin{equation}\label{QHLLE}
 Q^\kappa_{lr}(Q_l,Q_r)=\frac{b_r Q^\kappa_r-b_l Q^\kappa_l-F^\kappa(Q_r)+F^\kappa(Q_l)}{b_r-b_l}
\end{equation}
and the corresponding flux:
\begin{equation}\label{FHLLE}
 F^\kappa_{lr}(Q_l,Q_r)=\frac{b_r F^\kappa(Q_l)-b_l F^\kappa(Q_r)+b_l b_r(Q^\kappa_r-Q^\kappa_l)}{b_r-b_l},
\end{equation}
where $\kappa$ enumerates the Lorentz index and charge index.
For the completeness of the scheme one has to specify the signal velocities $b_l,b_r$. We take an advanced estimate for signal velocities from \cite{Rischke95}:
\begin{align}
b_r&=\max\left\{ 0, \frac{\bar v+\bar c_s}{1+\bar v\bar c_s}, \frac{v_r+c_{s,r}}{1+v_r c_{s,r}} \right\}, \\
b_l&=\min\left\{ 0, \frac{\bar v-\bar c_s}{1-\bar v\bar c_s}, \frac{v_l-c_{s,l}}{1-v_l c_{s,l}} \right\},
\end{align}
where $c_{s,r}=c_s(\epsilon_r)$, $c_{s,l}=c_s(\epsilon_l)$, and
\begin{align}
\bar v &= \frac{\sqrt{E_l}v_l+\sqrt{E_r}v_r}{\sqrt{E_l}+\sqrt{E_r}}, \\
\bar c_s^2 &= \frac{\sqrt{E_l}c_{s,l}^2+\sqrt{E_r}c_{s,r}^2}{\sqrt{E_l}+\sqrt{E_r}} + \eta\frac{\sqrt{E_l E_r}}{(\sqrt{E_l}+\sqrt{E_r})^2}(v_r-v_l)^2,
\end{align}
together with the suggested value of $\eta=0.5$.

For cells facing with vacuum, i.e. when $\epsilon_l=0$ or $\epsilon_r=0$, we put $b_l=-1$ or $b_r=1$ respectively.

For the second order accuracy of the scheme in space, a piecewise linear distributions in the cells (\texttt{MUSCL} scheme) are introduced. We reconstruct the values at right and  left cell boundary ($i\pm\frac 1 2$) as follows:
\begin{equation}
 Q_{i\pm}=Q_i\pm\frac 1 2 \Delta Q.
\end{equation}
Here we use the so-called \texttt{minmod} slope limiter:
$$\Delta Q=
\begin{cases}
\Delta_l,\quad\text{if}\ |\Delta_l|<|\Delta_r|\ \text{and}\ \Delta_l\cdot\Delta_r>0 \\
\Delta_r,\quad\text{if}\ |\Delta_l|>|\Delta_r|\ \text{and}\ \Delta_l\cdot\Delta_r>0 \\
0\quad\text{if}\ \Delta_l\cdot\Delta_r<0.
\end{cases}
$$
where $\Delta_l=Q_i-Q_{i-1}$, $\Delta_r=Q_{i+1}-Q_i$ are used. The slope limiter choses the smallest possible slope, and does not introduce new extrema. Therefore it avoids possible oscillations in the numerical solution.

To employ the piecewise linear distributions we substitute $Q_r\rightarrow Q_{(i+1)-}$, $Q_l\rightarrow Q_{i+}$ (and correspondingly $v_r\rightarrow v(Q_{(i+1)-})$, $v_l\rightarrow v(Q_{i+})$ etc.) in  Eq. \ref{QHLLE},\ref{FHLLE}.

For the second order accuracy in time we use the half-step (or predictor-corrector) method. First we propagate the evolution for half of timestep:
$$Q_i^{*n+\frac 1 2}=Q_i^n + \frac{\Delta t}{2\Delta x}(F^n_{i-\frac 1 2}-F^n_{i+\frac 1 2})+\frac{\Delta t}{2}S_i^n,$$
then the propagation is performed for a full timestep, based on fluxes and source terms calculated from $Q^{*n+\frac 1 2}$:
$$Q_i^{*n+1}=Q_i^n + \frac{\Delta t}{\Delta x}(F^{n+\frac 1 2}_{i-\frac 1 2}-F^{n+\frac 1 2}_{i+\frac 1 2})+\frac{\Delta t}{2}S_i^{n+\frac 1 2},$$
where $F^{n+\frac 1 2}=F(Q^{*n+\frac 1 2})$, $S^{n+\frac 1 2}=S(Q^{*n+\frac 1 2})$ and the propagated $Q$ is marked by an asterisk to keep the notation consistent with (\ref{num1ideal},\ref{num1visc}).

This completes the description of the scheme in one spatial dimension. To perform the evolution in three dimensions we apply the HLLE solver to calculate the fluxes through the cell boundaries independently for the $x$, $y$ and $\eta$ directions. The predictor step reads:
\begin{multline}
Q_{ijk}^{*n+\frac 1 2}=Q_{ijk}^n + \frac{\Delta t}{2\Delta x}(F^n_{i-\frac 1 2,jk}-F^n_{i+\frac 1 2,jk}) \\
+ \frac{\Delta t}{2\Delta y}(F^n_{i,j-\frac 1 2,k}-F^n_{i,j+\frac 1 2,k}) + \frac{\Delta t}{2\Delta \eta}(F^n_{ij,k-\frac 1 2}-F^n_{ij,k+\frac 1 2}) \\ + \frac{\Delta t}{2}S_{ijk}^n
\end{multline}
and the corrector step reads:
\begin{multline}
Q_{ijk}^{*n+1}=Q_{ijk}^n + \frac{\Delta t}{\Delta x}(F^{n+\frac 1 2}_{i-\frac 1 2,jk}-F^{n+\frac 1 2}_{i+\frac 1 2,jk})\\
 + \frac{\Delta t}{\Delta y}(F^{n+\frac 1 2}_{i,j-\frac 1 2,k}-F^{n+\frac 1 2}_{i,j+\frac 1 2,k})
 + \frac{\Delta t}{\Delta \eta}(F^{n+\frac 1 2}_{ij,k-\frac 1 2}-F^{n+\frac 1 2}_{ij,k+\frac 1 2})\\ +\frac{\Delta t}{2}S_{ijk}^{n+\frac 1 2}
\end{multline}
where the half-step updated variables are calculated from the full three dimensional predictor step.

It might happen that in a small amount of cells either $Q^{*n+1/2}_{ijk}$ or $Q^{*n+1}_{ijk}$ do not satisfy the requirement $Q^\tau>\sqrt{(Q^x)^2+(Q^y)^2+(Q^\eta)^2}$, which must hold since physical energy density is positive and the velocity $|\vec v|<1$. For these cases we proportionally rescale $\{Q^x,Q^y,Q^\eta\}$ wherever needed for the condition to be satisfied. This results in negligible deviations in a total energy-momentum balance in the system.

\subsection{Viscous substep}

In parallel to energy-momentum conservation equations, we integrate the equations of motion for the viscous terms, Eq. \ref{evolShearFinal},\ref{evolBulkFinal}.In the following we denote by $\pi$ either a component of $\pi^{\mu\nu}$ or $\Pi$, since the structure of their evolution equations is the same, except for different geometrical source terms. We split this substep into:

{\bf A)} Integration of source terms, which is performed using the predictor-corrector method:
\begin{align}
\pi^{\dagger n+1/2}&=\pi^{n}+I_\text{full}(\pi^n) \\
\pi^{\dagger n+1}&=\pi^{n}+I_\text{full}(\pi^{n+1/2})
\end{align}
where $I_\text{full}(\pi)=-(\pi-\pi_\text{NS})/\tau_\pi+I_\pi(\pi)$ as defined by (\ref{Ipi}).
Optionally, if $\tau_\pi$ is small enough, following the idea in \cite{TakamotoInutsuka} we use a formal solution to the equation with relaxation part only:
$$\pi^{\dagger n+1}=(\pi^n-\pi_\text{NS})\exp(-\frac{\Delta t}{\gamma\tau_\pi})+\pi_\text{NS},$$
and integrate $I_\pi$ separately. This is important since in heavy ion collision scenarios, depending on the ansatz and thermodynamical parameters taken,  $\tau_\pi$ may be comparable to the timestep.

{\bf B)} Advection using first order upwind method:
$$\pi^{n+1}_{ijk}=\sum_{\Delta i} \sum_{\Delta j} \sum_{\Delta k} w_{\Delta i}w_{\Delta j}w_{\Delta k}\pi^{\dagger n+1}_{i+\Delta i,j+\Delta j,k+\Delta k}$$
where $\Delta i, \Delta j, \Delta k=-1,0,+1$, and
$$w_{\Delta i}=\{-a_x^-, 1-|a_x|, a_x^+\}$$
$$a_x^-=\min(v_x \Delta t/\Delta x,0), \quad a_x^+=\max(v_x \Delta t/\Delta x,0)$$
with similar expressions for $a^\pm_y$ and $a^\pm_\eta$

The variables propagated for a half step are kept in memory and are used later for the calculation of the viscous fluxes and the source terms in the energy-momentum equations for the full timestep.

We evolve 10 independent components of $\pi^{\mu\nu}$, thus taking into account only that it is a symmetric tensor. This allows to check the consistency of the numerical solution by verification the achieved accuracy for the resulting orthogonality relations $\pi^{\mu\nu}u_\nu=0$ and the tracelessness relation $\pi^\mu_\mu=0$. An additional advantage is the simplicity of the velocity finding procedure, coming from the fact that one does not need to know the velocity to recover all components of $\pi^{\mu\nu}$.

It has been checked that employing the Lax-Wendroff method for advection substep and Strang splitting between advection/source substeps in heavy ion collision scenarios does not alter the evolution significantly. However, the upwind method is more stable for inhomogeneous distributions of $\pi^{\mu\nu}$ emerging from fluctuating initial conditions in event-by-event hydrodynamic simulations for heavy ion collisions.

The Israel-Stewart framework by itself does not restrict the values of the shear stress tensor or the bulk pressure. However, it is required that the viscous corrections are sufficiently small compared to the ideal quantities for the framework to be applicable. Nevertheless, in the practical applications it sometimes happens that $\pi^{\mu\nu}_\text{NS}$ or $\Pi_\text{NS}$ are not small due to large gradients of $u^\mu$ when the Lorentz-gamma factor is large. As a result, instabilities may develop in the hydrodynamical solution. To prevent this we monitor the conditions:
\begin{equation}
\max_{\mu,\nu}|\pi^{\mu\nu}|<C\cdot \max_{\mu,\nu}|T^{\mu\nu}_\text{id}|\quad\text{and}\quad |\Pi|<C\cdot p, \label{cutCond}
\end{equation}
where $C$ is some constant of the order one, but smaller than one. We rescale $\pi^{\mu\nu}$ and $\Pi$ where needed, to keep condition (\ref{cutCond}) satisfied on all hydro grid points. We found that condition (\ref{cutCond}) may only be violated in the regions with very small density during the matter expansion into the vacuum, as long as the initial conditions for dissipative quantities and values of relaxation times are within reasonable limits. In principle this indicates that in those regions the viscous hydrodynamic approximation becomes inapplicable. However, in heavy ion collision scenarios this does not affect the hydrodynamic evolution of the dense core region.

\subsection{Boundary conditions}
The cell average $Q^n_{ijk}$ is updated assuming that the values in neighbouring cells $Q^n_{i\pm2,j\pm2,k\pm2}$ are known. This is not the case for the cells on the boundary of hydrodynamic grid. Instead of introducing some special algorithm for them which depends on a type of boundary condition, we do somewhat easier procedure and extend the computational grid to include two additional cells on either end (in $x$,$y$,$\eta$ directions), called ghost cells. In applications to heavy ion collisions we study the matter expansion with vacuum. The computational boundary is therefore artificial and there should be no incoming signal, which means outflow (non-reflecting) boundary conditions. To realize it, at the beginning of each timestep the values of conservative variables in ghost cells are reset by the values from the nearest ``physical'' cell at either end of the grid, e.g. for $x$ direction:
$$Q^n_{N+2,jk}=Q^n_{N+1,jk}=Q^n_{N,jk},\quad Q^n_{0,jk}=Q^n_{1,jk}=Q^n_{2,jk},$$
where physical cells are in the range $[2,N]$. Then the fluxes are calculated between {\it all} cells which have both neighbours in a given direction.

\subsection{Final remarks}
As was mentioned above, the conserved quantities $Q^\alpha=\{T^{\tau\mu},N^\tau_c\}$, are used. For the completeness of the algorithm, one has to restore the so-called primitive variables - energy/charge densities and fluid velocity - several times during each timestep for each hydro cell: the fluxes/source terms have no explicit expressions in terms of $Q^\alpha$. Also the primitive variables are relevant for the output and further physical analysis. Obviously the recovery procedure should be fast. We employ a procedure, based on the one dimensional numerical root search as described in \ref{appendVelFind}.

\begin{figure*}
\includegraphics[width=0.49\textwidth]{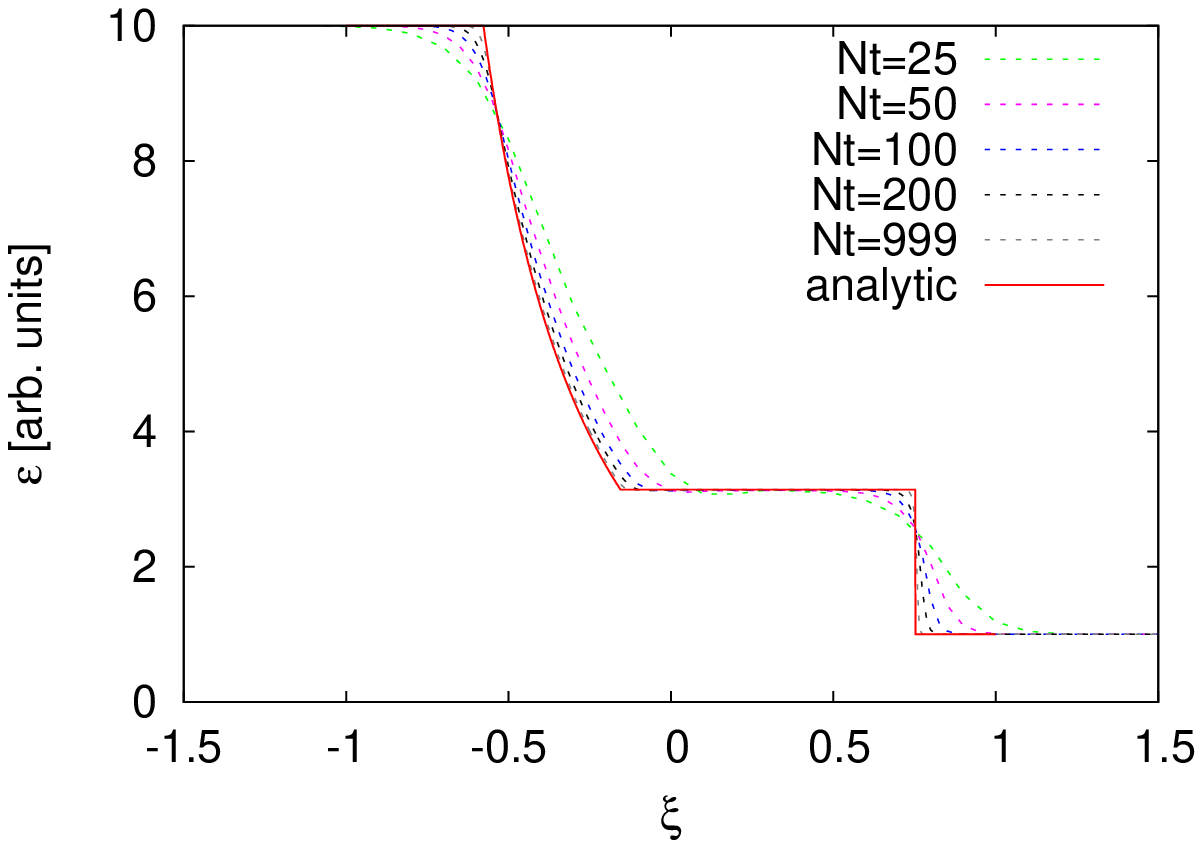}
\includegraphics[width=0.49\textwidth]{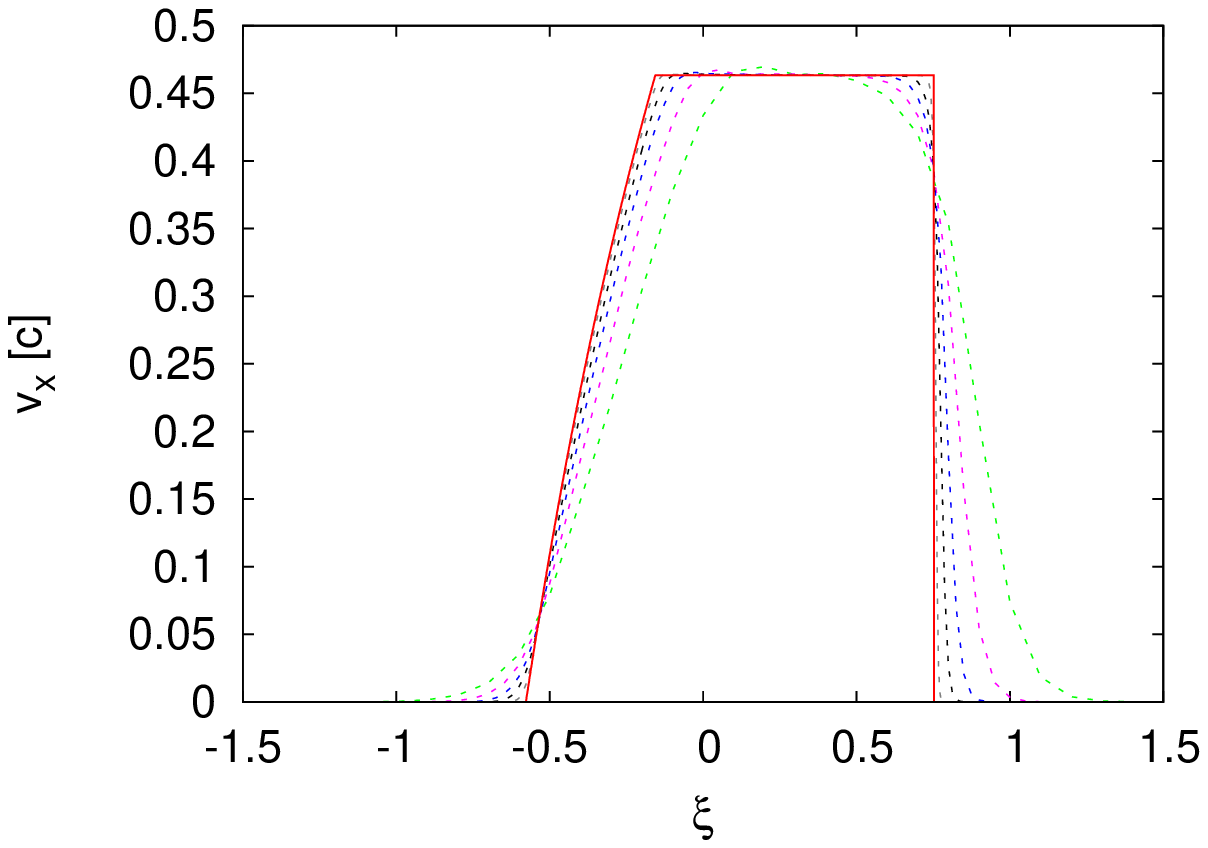}\\
\includegraphics[width=0.49\textwidth]{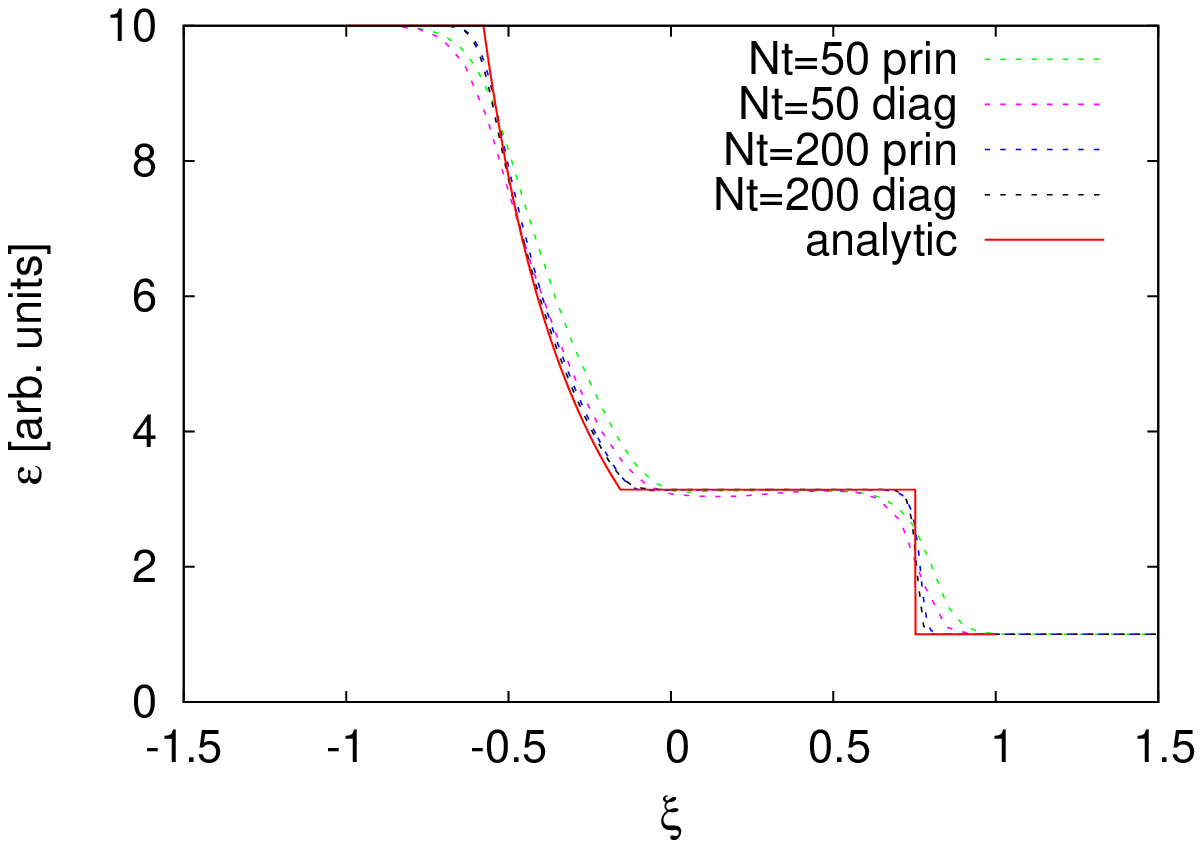}
\includegraphics[width=0.49\textwidth]{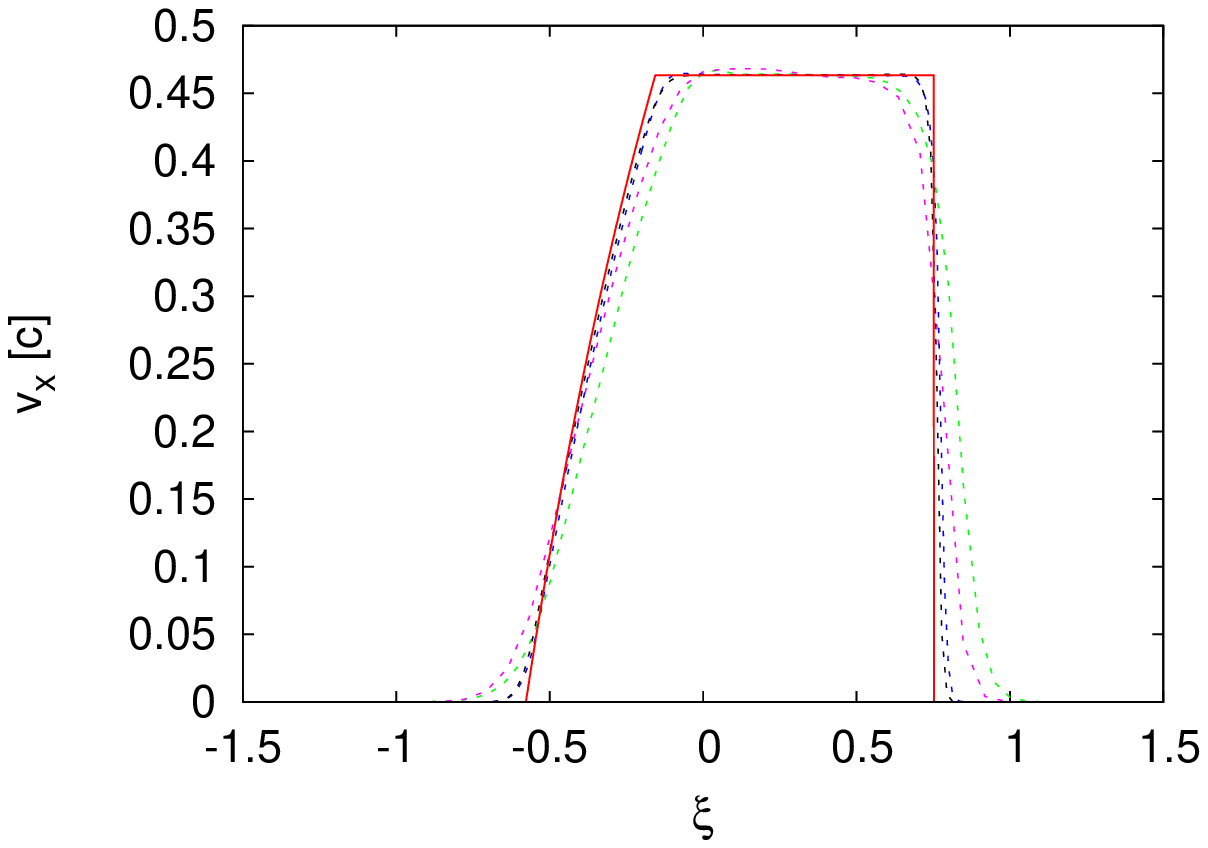}\\
\includegraphics[width=0.49\textwidth]{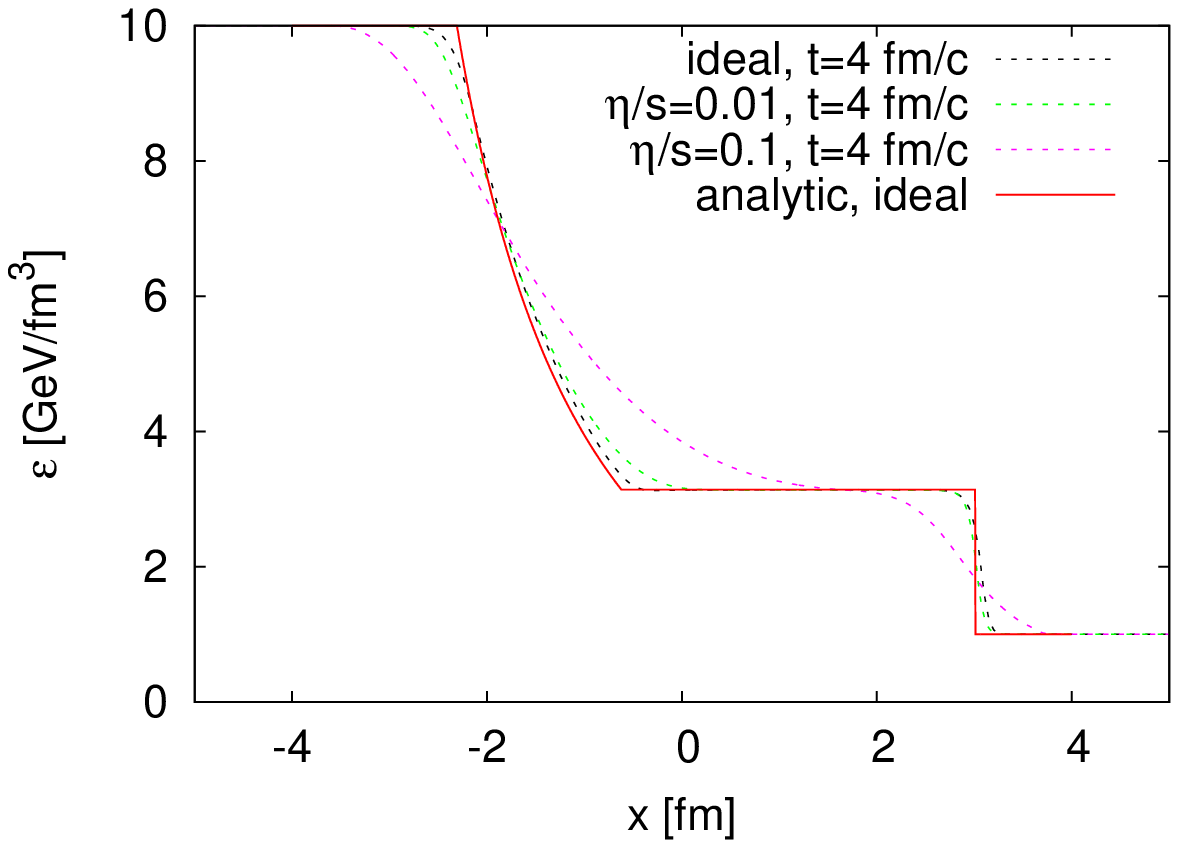}
\includegraphics[width=0.49\textwidth]{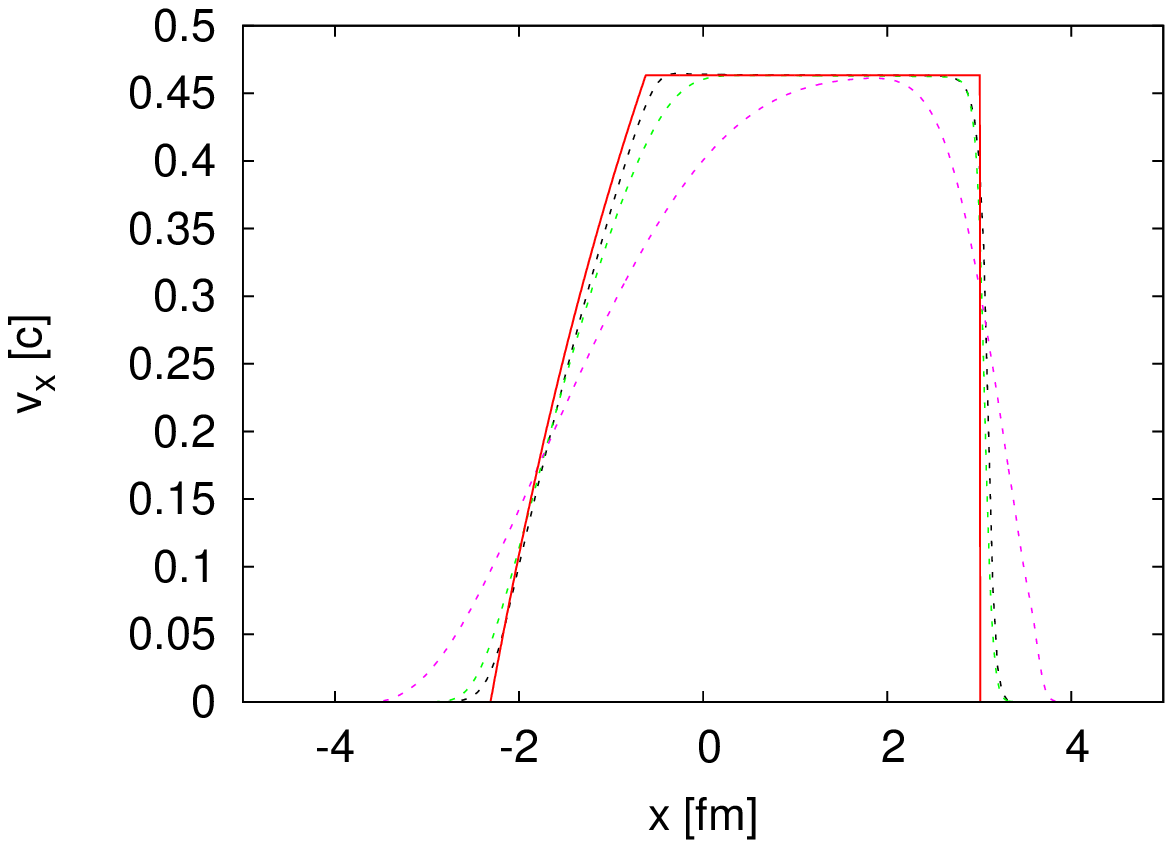}\\
\caption{(Color online) Analytical (solid line) and numerical (dashed lines) solutions to relativistic shock tube problem. Upper panels: comparison of numerical solution at different timesteps. Middle panels: comparison of the 1D numerical solution (denoted as \texttt{prin}, for principal direction in hydrodynamic grid) and 2D solution for 45 degree-rotated initial discontinuity (denoted as \texttt{diag}, for diagonal direction in hydrodynamic grid). Lower panels: ideal and viscous numerical solutions of shock tube problem. \texttt{Nt} denotes the number of timesteps $N_t$.}\label{figShockTube}
\end{figure*}

\section{Test results}\label{secTests}
\subsection{Ideal hydrodynamics: analytical solutions vs numerical solutions}
\textbf{Shock tube}. Let us start with a one dimensional shock tube problem. We initialize the system with two uniform (left and right) states $\{\epsilon_l=10\,\text{GeV/fm$^3$},v_l=0\}$ and $\{\epsilon_r=1\,\text{GeV/fm$^3$},v_r=0\}$, separated at $t<0$ by an imaginary membrane. The EoS for a relativistic massless gas $p=\epsilon/3$ is used. To extract the temperature or entropy density (for viscous hydro evolution) in this EoS we assume 2.5 massless quark degrees of freedom and $g_q=2\cdot2\cdot3=12$ degeneracy factor and $g_g=16$ for massless gluons.  At $t=0$ the membrane is removed and the initial discontinuity decays into compression shock wave propagating into the region of smaller density and a rarefaction wave propagating in the opposite direction. For such a case an analytical solution exists. The comparison between the analytical and numerical solution is shown in Fig. \ref{figShockTube}, upper panel. 
No scale parameters are present in such setup and the solution is expressed in terms of the dimensionless variable $\xi=x/t$. As it was pointed out in \cite{Rischke95}, it is essential to explore how many timesteps it takes for the numerical solution to approach the analytical one. The situation does not depend on the cell size provided that the Courant number $\lambda=\Delta t/\Delta x$ is kept the same. Since Eulerian grid is used, the wave profile is being resolved by the number of grid points/cells which increases with time (about 130 hydro cells at $N_t=200$). From the comparison one can see at the $N_t=25$th timestep there is substantial smearing of the profile, while at the $N_t=999$th timestep the profile is practically undistinguishable from the analytical result.

Next, to check the dependence of the simulations on the grid direction (rotational invariance) we rotate the initial discontinuity by 45 degrees in the $x-y$ plane and consider the same rarefraction/shock wave profile propagating in diagonal direction. The results are presented in the middle panel of Fig. \ref{figShockTube}. One can see that the propagation in the diagonal direction is consistent with principal direction at $Nt=200$, while there are some differences at $Nt=50$ when the numerical solution still does not approximate the analytical solution well.

Finally, the bottom panels of Fig. \ref{figShockTube} compare the simulations with shear viscosity to the solution of Riemann problem in the ideal case.

\begin{figure}
\includegraphics[width=0.49\textwidth]{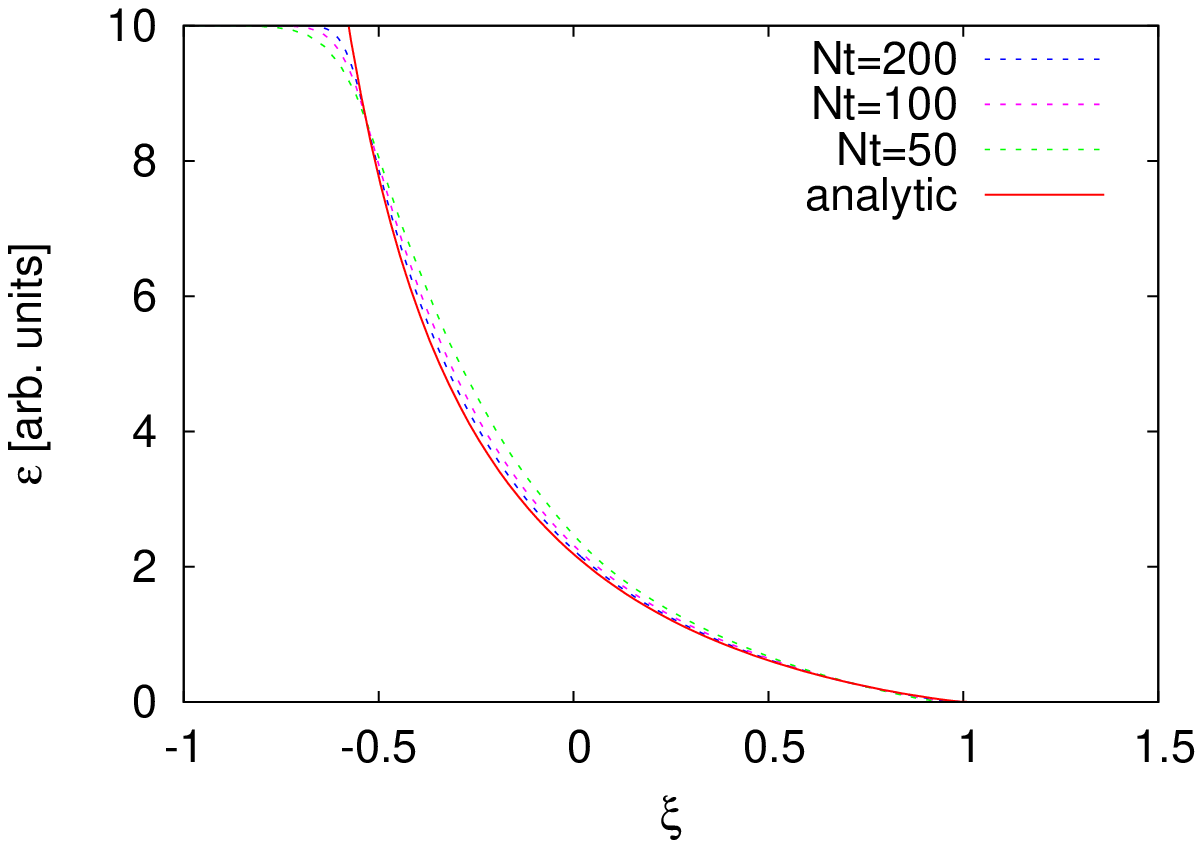}
\caption{(Color online) Energy density profile for analytical (solid line) and numerical (dashed lines) solutions to Riemann problem corresponding to matter expansion to vacuum}\label{figVacuum}
\end{figure}

\begin{figure}
\includegraphics[width=0.49\textwidth]{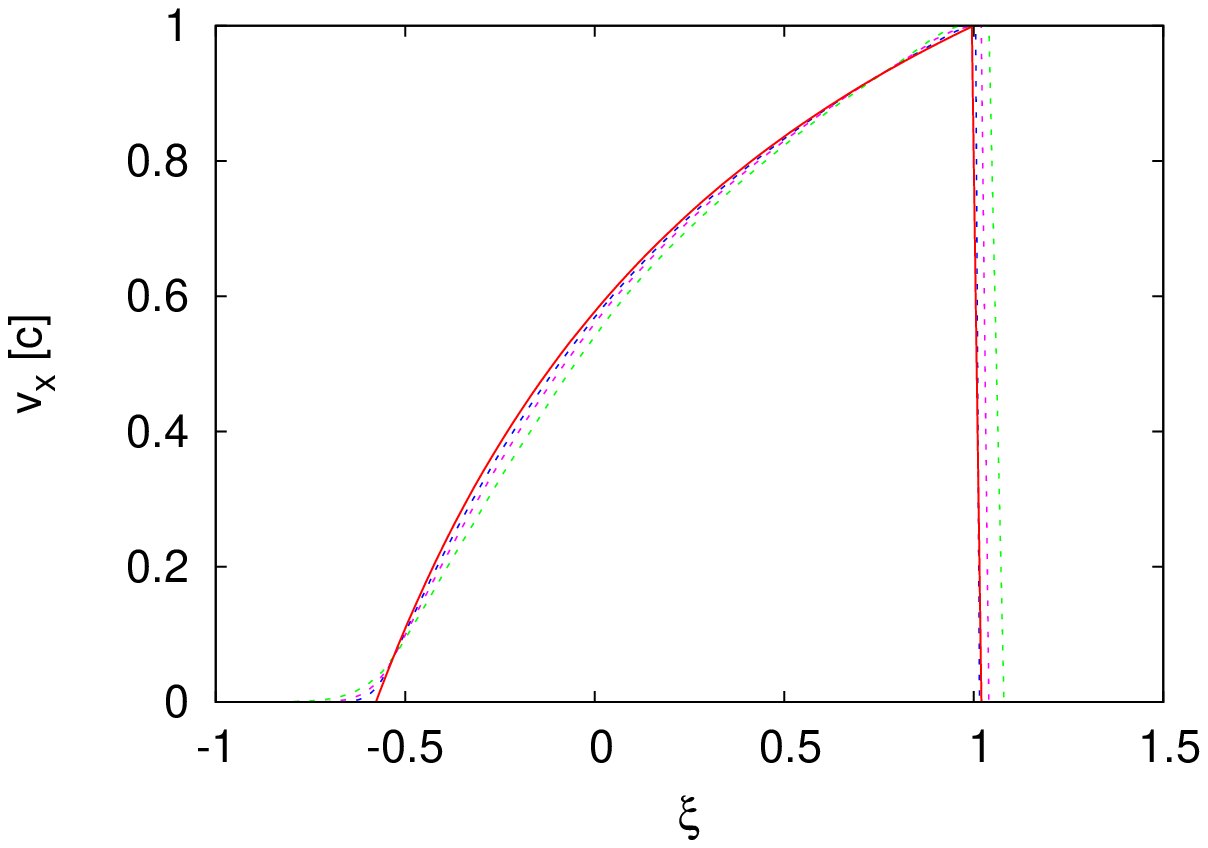}\\
\caption{(Color online) Same as Fig. \ref{figVacuum}, velocity profile.}\label{figVacuum2}
\end{figure}

In Figs. \ref{figVacuum}, \ref{figVacuum2} we consider a special case of the Riemann problem with $\{\epsilon_l=10\,\text{GeV/fm$^3$},v_l=0\}$ and $\{\epsilon_r=0,v_r=0\}$. This corresponds to a matter expansion to vacuum. In this case the analytical solution further depends on the dimensionless variable $\xi$, however at $t>0$ left and right (vacuum) states are connected with a rarefaction wave only, while the velocity of matter reaching the speed of light, $v=1$ at the boundary with vacuum. Fig. \ref{figVacuum} shows the results for the energy density profile (top) and the velocity profile (bottom). After $N_t=200$ timesteps the numerical solution approaches the analytical solution. % well. 
For this simulation we choose $\lambda_\text{CFL}=0.5$, thus the rarefaction wave at $N_t=200$ is spread over 100 hydrodynamic cells.

Since we take $\lambda_\text{CFL}<1$, one has to treat the rate of matter expansion to vacuum carefully. In the numerical solution, at each timestep matter from the boundary cells propagates to the next vacuum cells. This makes the effective velocity of the matter front to be $v_\text{front}=1/\lambda_\text{CFL}$, i.e. dependent on $\lambda_\text{CFL}$. We prevent this artefact by keeping the relative position of the matter front inside the cell, and allow to propagate to the next vacuum cell only after it crossed the current cell completely.

\textbf{Gubser flow}. Recently, a family of analytical relativistic hydrodynamic solutions was found for three-dimensional expansion of a conformal fluid, $p=\epsilon/3$. The solution assumes azimuthal symmetry in $xy$ plane and longitudinal scaling flow \cite{gubser}:
\begin{align}
 \epsilon &= \frac{\epsilon_0 (2q)^{8/3}}{\tau^{4/3}}\left[ 1+2q^2(\tau^2+r_T^2)+q^4(\tau^2-r_T^2)^2 \right]^{4/3},
\end{align}
\begin{align}
& u^\tau = \cosh[k(\tau,r_T)],  &  u^\eta = 0 \\
& u^x = \frac{x}{r_T}\sinh[k(\tau,r_T)],  &  u^y = \frac{y}{r_T}\sinh[k(\tau,r_T)]
\end{align}
\begin{equation}
k(\tau,r_T)=\text{arctanh}\frac{2q^2\tau r_T}{1+q^2\tau^2+q^2x_T^2}
\end{equation}
where $k(\tau,r_T)$ function has a meaning of transverse flow rapidity in Milne coordinates.

\begin{figure}
\includegraphics[width=0.49\textwidth]{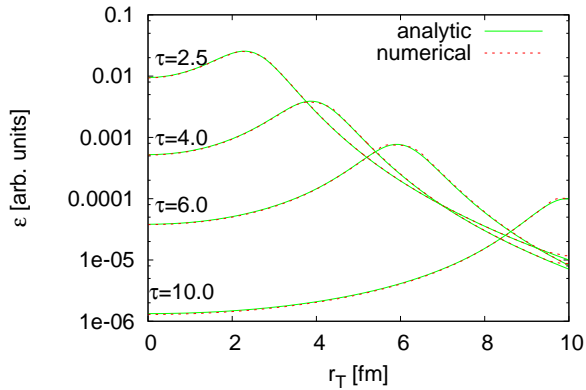}
\caption{(Color online) Energy density profile as a function of transverse coordinate at different times in analytical hydrodynamic solution by Gubser \cite{gubser}.}\label{figGubser}
\end{figure}

We set the parameters as follows: $\tau_0=1$ fm/c. $\epsilon_0=1$ [arbitrary~units], $q=1$ [arbitrary~units]. Fig. \ref{figGubser} depicts the comparison of the numerical solution to the analytical Gubser solution. The parameters correspond to an effective system size in transverse direction on the order of 1 fm, which is much smaller than the typical size of a heavy nucleus. Due to the strong initial transverse flow and persistent longitudinal flow the system expands and cools down very quickly. The evolution of this challenging initial state is reproduced by the numerical solution accurately, even after 10 fm/c time the very rarefied final state is reproduced well.

\begin{figure}
\includegraphics[width=0.49\textwidth]{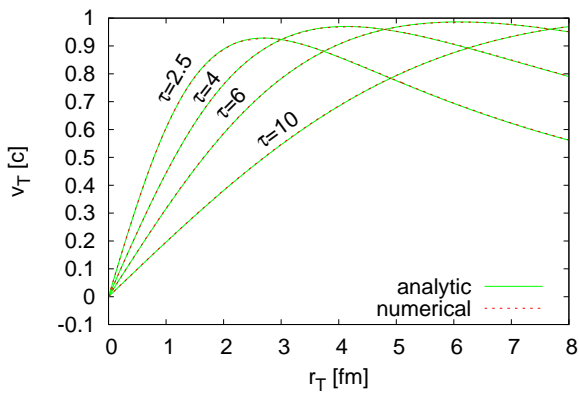}
\caption{(Color online) Same as Fig. \ref{figGubser}, transverse velocity profile}\label{figGubser2}
\end{figure}

\subsection{Viscous hydrodynamics: analytical solution vs. numerical solution} 
\begin{figure}
\includegraphics[width=0.49\textwidth]{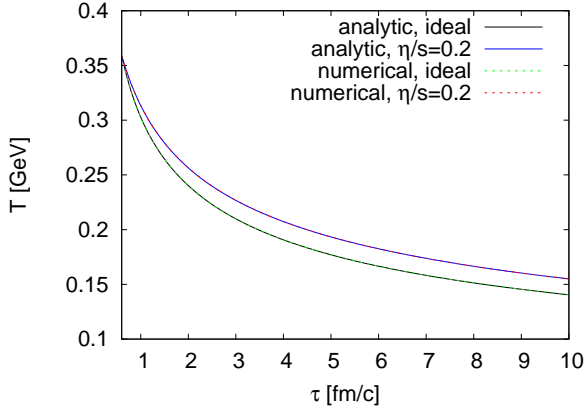}
\caption{(Color online) Analytical solutions for the evolution of temperature in 1D Bjorken expansion with and without shear viscosity (solid lines), compared to numerical solution (dashed lines).}\label{figAnalNS}
\end{figure}

It is also important to check the accuracy of the scheme in the viscous case.  Viscosity complicates the equations of relativistic hydrodynamics drastically. Thus analytical solutions exist only for very simple scenarios. First we consider the (0+1) dimensional Bjorken case. The system is homogeneous in all space directions with $v_x=v_y=v_\eta=0$ (which equals to scaling flow $v_z=z/t$) in the Navier-Stokes limit. Then $\pi^{\eta\eta}=-(4/3)\eta/\tau^3=-(4/3)(\eta/s)s/\tau^3$, and we obtain a modified Bjorken equation for the energy density evolution:
$$\frac{\dd\epsilon}{\dd\tau}+\frac{\epsilon+p+\tau^2 \pi^{\eta\eta}}{\tau}=0.$$
Assuming an ideal massless gas EoS, $p=\epsilon/3$ and $\epsilon=cT^4$, one obtains the analytical solution for $T(\tau)$ in the viscous case as
\begin{equation}
 T(\tau)=\left(\frac{\tau_0}{\tau}\right)^{1/3}\left[ T(\tau_0)+\frac{2\eta}{3s\tau_0}\left(1-\left(\frac{\tau_0}{\tau}\right)^{2/3}\right) \right].
\end{equation}
For the ideal fluid case $\eta/s=0$, the well known cooling law $T\propto \tau^{-1/3}$ for the scaling flow is restored. To compare to the numerical solution, the system is initialized with an energy density of $\epsilon_0=30$ GeV/fm$^3$ at $\tau_0=\sqrt{t^2-z^2}=0.6$ fm/c. Using the EoS for massless particles as described above, the initial temperature is $T_0=359$ MeV. We set $\eta/s=0.2$ and $\tau_\pi=0.0001$~fm (so that Navier-Stokes limit is well approximated) for the viscous case. Fig. \ref{figAnalNS} shows the comparison of the numerical solution for temperature to the analytical solution for the inviscid and viscous cases. One observes an agreement between the numerical and the analytical solutions.

\begin{figure}
\includegraphics[width=0.49\textwidth]{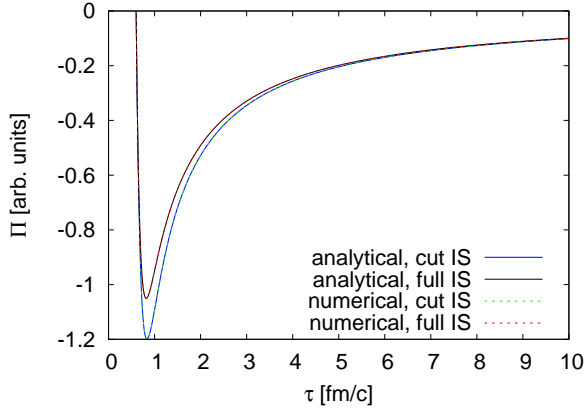}
\caption{(Color online) Evolution of bulk pressure $\Pi$ in analytical 0+1D solution to Israel-Stewart equations (solid lines) with 
the $(4\Pi)/(3\tau)$
%$\frac{4}{3}\Pi\dd_{;\mu}u^\mu$ 
term (indicated as ``full IS'') and without it (indicated as ``cut IS''). Dashed line is numerical solution.}\label{figAnalISBulk}
\end{figure}
To check how well the integration scheme for viscous fluxes works at finite $\tau_\pi$, we consider the evolution of the bulk pressure\footnote{The evolution equations for the non-trivial components of the shear stress tensor, $\pi^{xx}$, $\pi^{yy}$ or $\pi^{\eta\eta}$ are all very similar, therefore we only discuss the evolution of the bulk pressure.} in the same (0+1) dimensional case, but at finite $\tau_\Pi$. In this case, knowing that $\Pi_\text{NS}=\zeta/\tau$ one has:
\begin{equation}\label{bulkISanal}
 \frac{\dd\Pi}{\dd\tau}=-\frac{1}{\tau_\Pi}\left( \Pi-\frac{\zeta}{\tau} \right)-\frac{4\Pi}{3\tau}.
\end{equation}
Eq. \ref{bulkISanal}, \textit{without} the term $\frac{4\Pi}{3\tau}$ has an analytical solution expressed in terms of exponential integral function $\text{Ei}(x)$:
\begin{multline}
\Pi(\tau)=\Pi(\tau_0)e^{-(\tau-\tau_0)/\tau_\Pi}+\\+\frac{\zeta}{\tau_\Pi}e^{-\tau/\tau_\Pi}\left[ \text{Ei}(\tau_0/\tau_\Pi) - \text{Ei}(\tau/\tau_\Pi) \right]
\end{multline}
The inclusion of the $\frac{4\Pi}{3\tau}$ term leads to a more complicated analytical solution. In Fig. \ref{figAnalISBulk} a comparison between both (with and without the $\frac{4\Pi}{3\tau}$ term) analytical and numerical solutions is given. The specific set of parameters is $\zeta=1$, $\tau_\pi=0.1$~fm, $\tau_0=0.6$~fm and $\Pi(\tau_0)=0$.  One observes an excellent reproduction of both, the full and the ``cut'' (with and without $\frac{4\Pi}{3\tau}$ term, respectively) analytical viscous solutions.

% ------------ matter expansion in A+A -------------------------

\subsection{Matter expansion in heavy ion collisions}
Let us now turn to a more realistic scenario. We compare the present hydrodynamic simulations for a physical scenario related to heavy ion collisions with the open TECHQM results \cite{techqm}. The initial state has full homogeneity in $\eta$ direction ($v_\eta=0$, which corresponds to longitudinal scaling flow $v_z=z/t$) and the initial conditions in the transverse ($x-y$) plane are taken from the optical Glauber model for symmetric nucleus-nucleus collision:
\begin{multline}
\epsilon(\tau_0,r_x,r_y)=C\cdot n_{WN}(r_x,r_y)=\\
C\cdot T_A(r_x+\frac{b}{2},r_y)\left\{ 1-\left[1-T_A(r_x-\frac{b}{2},r_y)\frac{\sigma_{NN}}{A}\right]^A \right\}+\\
C\cdot T_A(r_x-\frac{b}{2},r_y)\left\{ 1-\left[1-T_A(r_x+\frac{b}{2},r_y)\frac{\sigma_{NN}}{A}\right]^A \right\}, \label{icGlauber}
\end{multline}
where the nuclear thickness function $T_A(x,y)=\int dr_z \rho(r_x,r_y,r_z)$ is normalized so that $\int T_A(x,y)dxdy=A$, and $\rho(r_x,r_y)=c/(exp[(r-R_A)/\delta]+1)$ is the density distribution for nucleons in the nucleus. For Au-Au collision the parameters are $A=197$, $R_A=6.37$~fm, $\delta=0.54$~fm, $\sigma_{NN}=40$~mb is the inelastic nucleon-nucleon cross section and $C$ is chosen so that $\epsilon_0(0,0;b=0)=30$~GeV/fm$^3$.

Again we use the EoS for a relativistic massless gas, $p=\epsilon/3$, assuming 2.5 massless quark degrees of freedom. The degeneracy factors are $g_q=2\cdot2\cdot3=12$ for quarks and $g_g=16$ for gluons.

For viscous hydrodynamic simulations the bulk viscosity is set to zero, $\pi^{\mu\nu}$ at $\tau_0$ is initialized with the Navier-Stokes values, yielding $\pi^{xx}=\pi^{yy}=-\tau^2\pi^{\eta\eta}/2=2\eta/(3\tau_0)$. The relaxation time for the shear is taken as $\tau_\pi=3\eta/(sT)$.

\begin{figure}
 \includegraphics[width=0.5\textwidth]{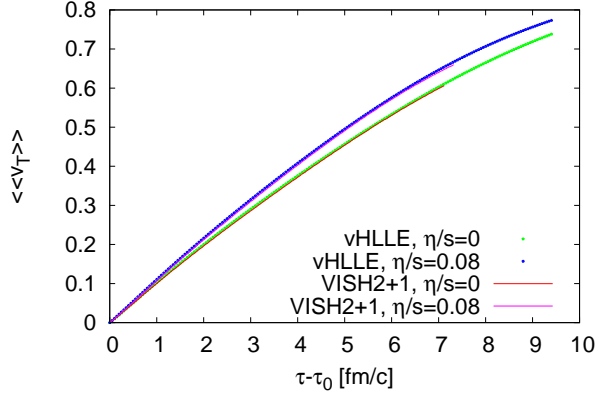}
 \caption{\label{figRadFlowSong}Averaged radial flow as a function of proper time for our hydro code (vHLLE) compared to \texttt{VISH2+1} \cite{VISH2p1}.}
\end{figure}

Fig. \ref{figRadFlowSong} shows the comparison between the present simulations and the (2+1) dimensional result by Song and Heinz for the average transverse velocity as a function of evolution time $\tau$ for initial conditions with impact parameter $b=0$. The average is defined as $$<<v_T>>=\int \frac{v_T\cdot\epsilon}{\sqrt{1-v_T^2}}d^2r_T$$
where $v_T=\sqrt{v_x^2+v_y^2}$ and the integration is made for a slice of the system (cells) with rapidity $y=0$. Shear viscosity works to equalize the expansion in different directions, thus decreasing work in longitudinal direction and accelerating the transverse expansion. This results in an additional acceleration of the transverse radial flow. Our results on the radial expansion for the ideal and viscous case are consistent with the benchmark results from the \texttt{VISH2+1} code.

\begin{figure}
 \includegraphics[width=0.5\textwidth]{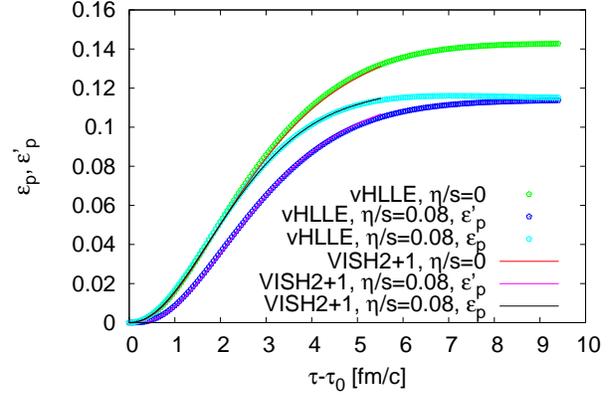}
 \caption{\label{figEpsilonpSong} Flow anisotropies $\epsilon_p$ and $\epsilon'_p$ (see text for explanation) as a function of proper time for our hydro code (vHLLE) compared to \texttt{VISH2+1} \cite{VISH2p1}.}
\end{figure}

In the same way shear viscosity suppresses the development of flow anisotropies in the transverse plane, the latter being generated by anisotropic pressure gradients in hydrodynamics. To explore this effect, we set the initial conditions to $b=7$~fm. Fig. \ref{figEpsilonpSong} shows the corresponding time evolution of the flow anisotropy, defined as
$$\epsilon_p=\frac{<T^{xx}_\text{id}-T^{yy}_\text{id}>}{<T^{xx}_\text{id}+T^{yy}_\text{id}>}$$
$$\epsilon'_p=\frac{<T^{xx}-T^{yy}>}{<T^{xx}+T^{yy}>}$$
where $<\dots>=\int\dots d^2r_T$. The quantities $\epsilon_p$ and $\epsilon'_p$ are calculated using the ideal part of the energy-momentum tensor and the full energy-momentum tensor, respectively. The observed suppression of $\epsilon_p$ in the viscous case relative to the ideal case comes solely from the rearrangement of collective flow, while $\epsilon'_p$ is suppressed stronger due to contributions from $\pi^{\mu\nu}$. The results are consistent with the benchmark results from the \texttt{VISH2+1} code \cite{VISH2p1}.

\begin{figure}
 \includegraphics[width=0.5\textwidth]{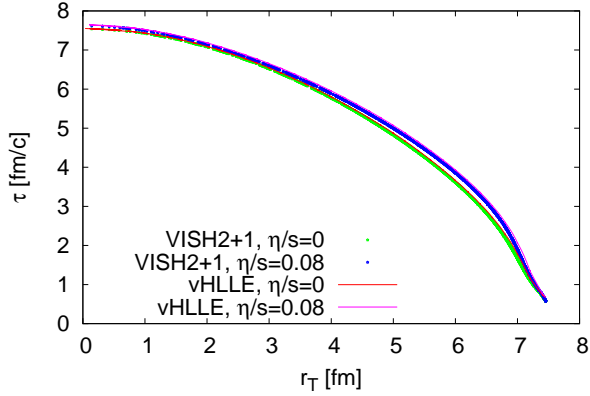}
 \caption{\label{figFreezeoutSong} Iso-thermal surface corresponding to $T_f=130$~MeV obtained with our hydro code (vHLLE) compared to the results  from \texttt{VISH2+1} \cite{VISH2p1}.}
\end{figure}

Finally, in Fig. \ref{figFreezeoutSong} we show the iso-thermal surfaces for the case $b=0$ corresponding to temperature $T_f=130$~MeV (or $\epsilon_f=0.516$~GeV/fm$^3$). The small differences (less than $~\Delta x/2=0.1$~fm) are related to the details (interpolation scheme) of the freezeout surface resolution.

\subsection{Energy conservation}
 The present scheme is conservative when Minkowski coordinates are used. However it loses the conservation property in Milne coordinates because the source terms are non-zero, and the accuracy of total energy conservation is determined by the source term integration part. To quantify the numerical accuracy of the energy conservation in a physical scenario, we run the code with initial conditions from the Glauber model (Eq. \ref{icGlauber}) with a limited rapidity profile, so that there is no energy/momentum leak through the grid edges in rapidity:
\begin{multline}\label{epsilon0_3d}
\epsilon(\tau_0,r_x,r_y,\eta_s)=C N_\text{WN} \theta(Y_b-\eta_s)\cdot \\
\cdot\exp\left[ -\theta(|\eta_s|-\Delta\eta)\frac{(|\eta_s|-\Delta\eta)^2}{\sigma_\eta^2} \right]
\end{multline}
where $Y_b=5.3$ corresponds to the beam rapidity, $\sigma_\eta=2.1$, and $\Delta\eta=1.3$ is the size of plateau around midrapidity. The hydrodynamic grid consists of $n_x\cdot n_y\cdot n_z=150\cdot150\cdot100$ cells with $\Delta x=\Delta y=0.2$~fm/c, $\Delta\eta=0.2$ units, and corresponding $\Delta\tau=0.05$.
The total energy on the hypersurface of constant $\tau$ is defined as $E_\text{tot}(\tau)=\int T^{0i}d\sigma_i$, which can be expanded as:
\begin{multline}
E_\text{tot}=\tau\int d\eta d^2r_T [(\epsilon+p)\tilde u^\tau(\tilde u^\tau \cosh\eta+\tilde u^\eta \sinh\eta)-\\-p \cosh\eta+\tilde \pi^{\tau\tau}\cosh\eta+\tilde \pi^{\tau\eta}\sinh\eta]
\end{multline}
\begin{equation}
 S_\text{tot}=\tau\int d\eta d^2r_T \cdot s\tilde u^\tau
\end{equation}

 Numerically $\int d\eta d^2r_T(...)\rightarrow \Delta x \Delta y \Delta\eta \sum\limits_\text{cells}(...)$.

\begin{table}[h]
\begin{tabular}{|c|c|c|c|}
\hline
 $\eta/s$ & 0 & 0 & 0.1 \\ \hline
 EoS & $p=\frac{\epsilon}{3}$ & Laine \cite{Laine} & $p=\frac{\epsilon}{3}$ \\ \hline\hline
 $E_\text{tot}(\tau=1)$ [GeV] & 68230 & 68230 & 68230 \\ \hline
 $E_\text{tot}(\tau=10)$ [GeV] & 69419  & 69537  & 69927 \\
 \ & {\tiny (+1.7\%)} & {\tiny(+1.9\%)} & {\tiny(+2.5\%)} \\ \hline\hline
 $S_\text{tot}(\tau=1)$ & 37734 & 45469 & 37734 \\ \hline
 $S_\text{tot}(\tau=10)$ & 37884  & 45632  & 40442 \\
 \ & {\tiny(+0.4\%)} & {\tiny(+0.35\%)} & {\tiny(+6.9\%)} \\ \hline
\end{tabular}
\caption{Total energy and entropy calculated in the beginning [$\tau=1$ fm/c] and in the end [$\tau=10$ fm/c] of 3D hydrodynamic evolution with initial energy density profile (\ref{epsilon0_3d}) for different viscosity/EoS combinations. Small numbers in parentheses denote percentage of increase compared to the value at $\tau=1$ fm/c.}\label{tbEtot}
\end{table}

It is important to note that in Israel-Stewart framework the entropy current $s^\mu$ includes non-equilibrium corrections:
$$s^\mu=s_\text{eq}-(\frac{\beta_0}{2T}\Pi^2+\frac{3}{2(\epsilon+p)T}\pi^{\mu\nu}\pi_{\mu\nu})u^\mu$$
where the coefficient in front of $\pi^{\mu\nu}\pi_{\mu\nu}$ is taken consistently with the evolution equations (\ref{Ipi}), and $\Pi=0$ since we consider shear viscosity only.

 The resulting values of total energy and entropy in the beginning and in the end of hydrodynamic evolution are shown in Table \ref{tbEtot}. We conclude that energy is conserved on a level better than 3\%.

%-------------- numerical viscosity ---------------

\subsection{Numerical viscosity}
Since we study the effects of physical viscosity with the code, the important question which has to be answered is: what amount of numerical viscosity the code has, and how does it depend on the parameters?

To study this, we follow the method used in \cite{NonakaTI} and examine the sound wave attenuation in numerical hydrodynamic solution. The initial conditions for 1D hydrodynamic simulation in Minkowski coordinates are taken as
\begin{align}
\epsilon(x)&=\epsilon_0+\delta\epsilon\sin(2\pi x/\lambda), \\
v_x(x)&=\frac{c_s\delta\epsilon}{\epsilon_0+p_0}\sin(2\pi x/\lambda),
\end{align}
supplemented by an EoS for an ultrarelativistic gas, $p=\epsilon/3$. Given that $\delta\epsilon\ll\epsilon_0$, this represents a sound wave with length $\lambda$ propagating on the static uniform background with energy density $\epsilon_0$.
 We link the last cell in $x$ direction to the first one to set up periodic boundary conditions on the hydro mesh, so that $\epsilon(-\lambda/2)=\epsilon(\lambda/2)$, $v_x(-\lambda/2)=v_x(\lambda/2)$. This setup mimics the propagation of a plain sound wave over an infinite medium. In hydrodynamics the attenuation (or damping) of the sound wave amplitude is only possible due to viscosity. Provided that the damping is not fast (i.e. that the amplitude does not change significantly during one cycle) there is an analytical expression for the amplitude of sound wave after one cycle $t=\lambda/c_s$:
\begin{equation}
\delta\epsilon(\lambda/c_s;\eta)=\delta\epsilon(0;\eta)\exp\left( -\frac{8\pi^2\eta}{3\lambda c_s(\epsilon_0+p_0)} \right),
\end{equation}
here and further $c_s=c_s(\epsilon_0)$. However, due to the presence of numerical dissipation, the attenuation of the sound wave is also possible even with zero physical viscosity. To quantify the differences between the solutions we calculate the L1 norm for the energy density function on one-wavelength segment, defined as:
$$L(N_\text{cell})=\frac{\lambda}{N_\text{cell}}\sum_{i=1}^{N_\text{cell}}\left|\epsilon_\text{num}(x_i,\lambda/c_s)-\epsilon_\text{anal}(x_i,\lambda/c_s)\right|$$
which is compared to the same quantity based on the difference between inviscid and viscous solutions:
$$L_\text{phys}(\eta)=\frac{\lambda}{N_\text{cell}}\sum_{i=1}^{N_\text{cell}}\left|\epsilon_\text{anal}(x_i,\frac{\lambda}{c_s};\eta)-\epsilon_\text{anal}(x_i,\frac{\lambda}{c_s};0)\right|$$
For the latter quantity, in the limit of an infinitely small cell size one gets an analytical result:
$$L_\text{phys}(\eta)=\frac{2\lambda}{\pi}\delta\epsilon(0)\left(1-\exp\left( -\frac{8\pi^2\eta}{3\lambda c_s(\epsilon_0+p_0)} \right)\right)$$
which is also quite accurate for finite cell size. Comparing the two quantities, $L_\text{phys}(\eta_\text{num})=L(N_\text{cell})$ one gets:
\begin{equation}
\eta_\text{num}=-\frac{3\lambda}{8\pi^2}c_{s}(\epsilon_0+p_0)\ln\left[1-\frac{\pi}{2\lambda\delta\epsilon(0)}L(N_\text{cell})\right]
\end{equation}
\begin{figure}
 \includegraphics[width=0.5\textwidth]{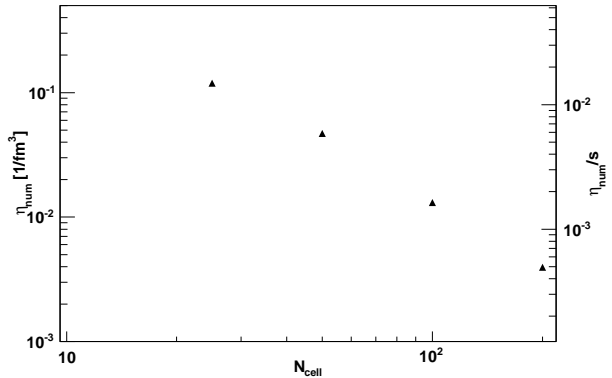}\caption{Numerical viscosity $\eta_\text{num}$ and corresponding $\eta_\text{num}/s$ values (the latter using upper estimate for the temperature $T=0.5$~GeV) observed from sound wave attenuation for different grid sizes.}\label{figEtaNum}
\end{figure}
We initialize the system with the following parameter values: $\lambda=10$ fm, $\epsilon_0=3$ GeV/fm$^3$ and $\delta\epsilon(0)=0.003$ GeV/fm$^3$. The values of the resulting numerical viscosity for different grid sizes are shown in Fig. \ref{figEtaNum}. Since the shear viscosity coefficient is proportional to density, the relevant (and dimensionless) quantity for relativistic case is the ratio of shear viscosity to entropy density, $\eta/s$. Assuming zero chemical potential, one gets:
\begin{equation}
\frac{\eta_\text{num}}{s}=-\frac{3\lambda T}{8\pi^2}c_{s}\ln\left[1-\frac{\pi}{2\lambda\delta\epsilon(0)}L(N_\text{cell})\right].
\end{equation}
Note that since $L(N_\text{cell})\propto\epsilon_0$, the above expression depends only on $\lambda$ and $T$. Assuming $T=0.5$~GeV as an upper estimate for the initial phase of the hydrodynamic expansion in A+A collisions and $\lambda=10$~fm, one gets $\eta_\text{num}/s=0.015$ for $N_\text{cell}=25$. For $N_\text{cell}=100$, which is a typical grid size for the physical simulations, $\eta_\text{num}/s=0.0016$. I.e., the value of the numerical viscosity to entropy density is about 50 times smaller than the lower bound for the physical viscosity $(\eta/s)_\text{min}=1/4\pi$.

We should finally note that the estimate does not guarantee that a similar amount of numerical viscosity is present in full-fledged (3+1) dimensional simulations of matter expansion with the given code. To estimate the numerical viscosity in arbitrary geometry is a rather complicated topic beyond this paper.

\section{Conclusions}\label{secConclusions}
We have presented a detailed description and test results of a (3+1) dimensional relativistic viscous hydrodynamic code based on the Godunov method and the relativistic HLLE approximation for the solution of the Riemann problem for its inviscid part. This choice ensures that the code is capable of treating shock wave configurations accurately. It has been shown that the code is capable of solving the equations of relativistic viscous hydrodynamics in the Israel-Stewart framework with the help of the ideal-viscous splitting method. We have presented the results of several test problems: the 1 dimensional (2 dimensional) shock tube, Gubser flow and two analytical viscous hydrodynamic solutions. The numerical viscosity of the code in the inviscid case has been estimated and found to be sufficiently small.

The primary application of the code is the simulations of the hydrodynamic expansion of QCD matter created in relativistic heavy ion collisions. For this aim we have also checked the code against the test cases by the TECHQM group.

%--------------- appendix ------------------

\section{Acknowledgements}
The authors acknowledge the financial support by the ExtreMe Matter Institute EMMI and Hessian LOEWE initiative. The work of P.H. was supported by BMBF under contract no. 06FY9092. Computational resources have been provided by the Center for Scientific Computing (CSC) at the Goethe-University Frankfurt.

\appendix
\section{Velocity finding}\label{appendVelFind}
An important part of hydrodynamic algorithm is the procedure to find the flow velocity $v_i$, energy density $\epsilon$ and densities of conserved charges $n_k$ in the fluid rest frame from the conserved variables $T^{\tau\mu},N^\mu$.

The procedure is essentially the same for inviscid and viscous cases. In the latter case to account for shear viscosity, one has to subtract the $\pi^{\tau\mu}$ (which are evolved independently with IS equations) from the total energy-momentum tensor: $T^{\tau\mu}_\text{id}=T^{\tau\mu}-\pi^{\tau\mu}$.

The definition of the energy-momentum tensor of the fluid gives the following system of equations:
\begin{align}
T^{\tau\tau}_\text{id}&=E=(\epsilon+p)/(1-v^2)-p, \nonumber\\
T^{\tau x}_\text{id}&=M_x=(\epsilon+p)v_x/(1-v^2), \nonumber\\
T^{\tau y}_\text{id}&=M_y=(\epsilon+p)v_y/(1-v^2), \nonumber\\
T^{\tau \eta}_\text{id}&=M_\eta=(\epsilon+p)v_\eta/(1-v^2), \nonumber\\
N_c&=n_c/\sqrt{1-v^2} \nonumber\\
\end{align}
in terms of $\epsilon$, $v_x$, $v_y$, $v_\eta$ and $n_c$ ($c$ is numbering the conserved charges), which is closed with an equation of state:
$$p=p(\epsilon,n_i)$$
Due to the symmetry of the equations, $v_i(E+p)=M_i$. This allows one to reduce the problem to a one-dimensional equation for the absolute value of the velocity, which has to be solved numerically \cite{Rischke95}:
\begin{equation}\label{velocityRoot}
v=\frac{|\vec M|}{E+p(E-\vec{M}\cdot\vec{v},N_i\sqrt{1-v^2})}
\end{equation}
then the rest of unknowns are recovered as
\begin{align}
& v_x=M_x/|\vec M|  &  \epsilon=E-\vec{M}\cdot\vec{v}  \nonumber\\
& v_y=M_y/|\vec M|  &  n_i=N_i\sqrt{1-v^2}  \nonumber\\
& v_\eta=M_\eta/|\vec M|  &  \nonumber
\end{align}
For non-exotic equation of state Eq. (\ref{velocityRoot}) has exactly one root in the interval $v=[0\dots1)$ and is solved with Newton's method.

In the presence of bulk pressure one has to add it to the equilibrium pressure, $p(\epsilon,n_i)\rightarrow p(\epsilon,n_i)+\Pi$ and proceed to solve (\ref{velocityRoot}) in the same way.

\end{document}